\documentclass[lettersize,journal]{IEEEtran}
\usepackage[colorinlistoftodos, textsize=small]{todonotes}
\usepackage{amsmath,amsfonts}
\usepackage{soul}
\usepackage{algorithm,algorithmic}
\usepackage{array}
\usepackage{caption}
\usepackage{tikz}
\usepackage[colorlinks,
            linkcolor=blue,       
            anchorcolor=blue,  
            citecolor=blue,        
            ]{hyperref}
\usepackage{subfig}

\usepackage{textcomp}
\usepackage{stfloats}
\usepackage{url}
\usepackage{verbatim}
\usepackage{graphicx}
\usepackage[table,xcdraw]{xcolor}
\usepackage{cite}
\usepackage{amssymb}
\usepackage{booktabs} 
\usepackage{balance}
\usepackage{multirow}
\usepackage[normalem]{ulem}
\usepackage{colortbl}   

\definecolor{myblue}{RGB}{235, 245, 255}

\useunder{\uline}{\ul}{}
\def\BibTeX{{\rm B\kern-.05em{\sc i\kern-.025em b}\kern-.08em
    T\kern-.1667em\lower.7ex\hbox{E}\kern-.125emX}}
\begin{document}
\title{A H.265/HEVC Fine-Grained ROI Video Encryption Algorithm Based on Coding Unit and Prompt Segmentation}
\author{Xiang Zhang, Haoyan Lu, Ziqiang Li, Ziwen He, Zhenshan Tan, Fei Peng, Zhangjie Fu

\thanks{This work was supported in part by the National Natural Science Foundation of China under Grant 62202234, 62372128, 62401270, U22B2062, 62172232; China Postdoctoral Science Foundation under Grant 2023M741778; Natural Science Foundation of Guangdong Province under Grant 2023A1515011575; Nanjing Major Science and Technology Special Project under Grant 202405002.

Xiang Zhang, Haoyan Lu, Ziqiang Li, Ziwen He, Wenbin Huang, and Zhangjie Fu are with the Engineering Research Center of Digital Forensics, Ministry of Education, Nanjing University of Information Science and Technology, Nanjing, Jiangsu 210044, China (e-mail: zhangxiang@nuist.edu.cn; 202312490279@nuist.edu.cn; fzj@nuist.edu.cn).

Fei Peng is with the School of Artificial Intelligence, Guangzhou University, Guangzhou, Guangdong 510006, China (e-mail: eepengf@gmail.com).
}}

\markboth{Journal of \LaTeX\ Class Files,~Vol.~14, No.~8, August~2021}%
{Shell \MakeLowercase{\textit{et al.}}: A Sample Article Using IEEEtran.cls for IEEE Journals}


\maketitle

\begin{abstract}
ROI (Region of Interest) video selective encryption based on H.265/\ HEVC is a technology that protects the sensitive regions of videos by perturbing the syntax elements associated with target areas. However, existing methods typically adopt Tile (with a relatively large size) as the minimum encryption unit, which suffers from problems such as inaccurate encryption regions and low encryption precision. This low-precision encryption makes them difficult to apply in sensitive fields such as medicine, military, and remote sensing. In order to address the aforementioned problem, this paper proposes a fine-grained ROI video selective encryption algorithm based on Coding Units (CUs) and prompt segmentation. First, to achieve a more precise ROI acquisition, we present a novel ROI mapping approach based on prompt segmentation. This approach enables precise mapping of ROIs to small $8\times8$ CU levels, significantly enhancing the precision of encrypted regions. Second, we propose a selective encryption scheme based on multiple syntax elements, which distorts syntax elements within high-precision ROI to effectively safeguard ROI security. Finally, we design a diffusion isolation based on Pulse Code Modulation (PCM) mode and MV restriction, applying PCM mode and MV restriction strategy to the affected CU to address encryption diffusion during prediction. The above three strategies break the inherent mechanism of using Tiles in existing ROI encryption and push the fine-grained level of ROI video encryption to the minimum $8\times8$ CU precision. The experimental results demonstrate that the proposed algorithm can accurately segment ROI regions, effectively perturb pixels within these regions, and eliminate the diffusion artifacts introduced by encryption. The method exhibits great potential for application in medical imaging, military surveillance, and remote areas.
\end{abstract}


\section{Introduction}
\IEEEPARstart{A}{s} a powerful information medium, digital video has permeated all aspects of contemporary society, finding extensive applications in diverse scenarios such as social media, telecommunication, intelligent surveillance systems, and medical detection. With the rapid development of digital media technology, people have put forward higher demands for the capabilities of video data processing and transmission. For instance, in the field of telecommunication, video data, as an important information carrier, can provide people with real-time "face-to-face" communication support. However, it inherently requires capturing a large amount of sensitive information, especially biometric identifiers such as facial features. Such practices are bound to lead to privacy and security issues such as unauthorized access and potential identity leakage during the transmission and model training of video data\cite{PADILLALOPEZ20154177}.

Video encryption is one of the important technical means for privacy protection. Early naive encryption methods perform full encryption on the original video bitstream, which neglects key factors such as coding structure, compression efficiency, and security policies. Although such methods can effectively protect data, they are not compatible with video coding technologies and are therefore difficult to widely apply \cite{5955103}. In contrast, selective encryption methods have emerged as the times require due to their compatibility with video coding technologies. Their core idea is to encrypt only specific syntax elements in the video bitstream, thus significantly reducing the computational overhead. Among numerous video coding standards, H.265/HEVC is currently one of the most mainstream video coding standards \cite{6626250}, so selective encryption algorithms based on H.265/HEVC have also become a research hotspot in the field of video encryption \cite{6316136,zhang2025visual}.

\begin{figure}[t]
    \centering
    \includegraphics[width=1\linewidth]{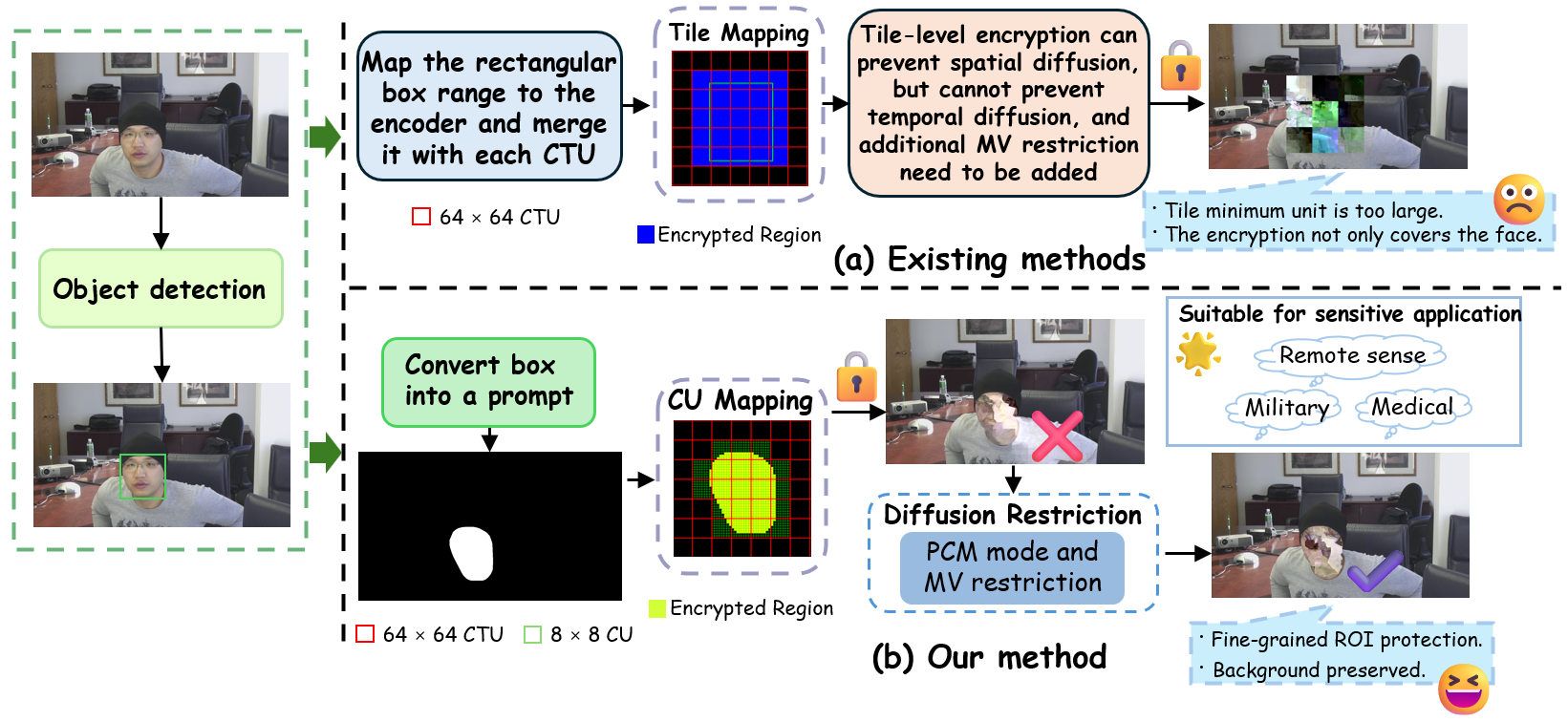}
    \caption{Comparison between the existing methods and the proposed method. The existing methods encrypt coarse Tile-aligned regions, which may cover non-ROI, while the proposed method achieves finer ROI localization through CU-level mapping, combines PCM mode, MV restriction, and CU-level encryption to reduce distortion leakage.}
    \label{fig:moti}
\end{figure}

Selective encryption schemes based on H.265/HEVC can be divided into two categories: global encryption and Region of Interest (ROI) encryption \cite{dufaux2008scrambling,dufaux2008h,carrillo2008compression,unterweger2016building,zhang2018lightweight,hosny2022privacy,cho2024practical,farajallah2015roi,tew2016region,taha2018end,yu2023coding,li2024ppl,zhang2025visual}. The former encrypts all regions of video frames without considering video semantic information, while the latter only encrypts designated regions (such as human faces, license plates, or specific objects), aiming to protect the privacy of the ROI while preserving the visibility of the non-ROI background regions \cite{farajallah2015roi}. At present, existing ROI encryption methods mainly focuses on Tile-level encryption \cite{misra2013overview}. As can be seen from Figure \ref{fig:moti}(a), once the ROI is positioned, encryption is usually performed on coarse Tile-level coding blocks. Due to insufficient precision, the background regions are inevitably disturbed simultaneously, which not only reduces the accuracy of encryption but also impairs the visibility of non-ROI. In many conventional video application scenarios, The encryption overflow into the background regions may not pose a significant problem. However, in sensitive domains such as medicine, military reconnaissance, and remote sensing, it is often necessary to protect ROI—for example, identity-related areas—while still preserving background information required for critical analysis tasks, such as lesion localization. Any distortion or interference introduced into background regions can severely compromise analytical accuracy and decision-making reliability. Consequently, Tile-level ROI encryption strategies, which inherently operate at coarse granularity, are ill-suited for such applications.

To overcome the limitations imposed by the existing Tile mechanism and to substantially enhance the granularity of ROI protection, we propose a fine-grained ROI selective encryption algorithm based on coding unit and prompt segmentation, as shown in Figure \ref{fig:moti}(b). Our proposed scheme includes a ROI mapping method based on prompt segmentation. This method combines object detection with prompt segmentation to obtain a pixel-level ROI segmentation mask, which is then mapped to the $8\times8$ CU level through a specially designed mask mapping rule. Moreover, although CU-level encryption enables more precise coverage of object regions, significant encryption diffusion occurs in most cases. This phenomenon stems from the strong spatial and temporal dependencies inherent in the H.265/HEVC prediction process \cite{sze2014high}, where perturbations of the encrypted CUs will spread to adjacent non-ROI. Existing ROI encryption methods naturally avoid this issue by exploiting Tile-level independence (Tiles do not reference one another). However, when the encryption precision is increased to the $8\times8$ CU level without using the Tile mechanism, encryption diffusion will be inevitable, posing a significant challenge to high-precision ROI protection. To address this problem, We further propose a diffusion isolation mechanism that combines the PCM mode with MV restriction. In summary, the main contributions of this paper are as follows:

\begin{itemize}
\item \textbf{A high-precision ROI mapping framework based on prompt segmentation.} The framework first introduces object detection to localize the ROI regions. Then, it utilizes prompt segmentation to generate pixel-level masks that closely match the true contours. Finally, these masks are directly mapped to the adaptive fine-gained CUs of H.265/HEVC by our proposed mapping strategy. This framework breaks through the limitations of Tile-level ROI encryption and fundamentally solves the problems of over-encryption caused by inaccurate region delimitation in the existing methods.

\item \textbf{A selective encryption scheme based on multiple syntax elements.} We propose an encryption scheme based on multiple syntax elements, which combines regular coding and bypass coding to encrypt several syntax elements, including IPM, MVPIdx, MergeIdx, RefFrmIdx, the values and signs of MVD, as well as residual coefficients. This strategy effectively enhances the security of encrypted regions and ensures the protection of privacy information.

\item \textbf{A diffusion isolation mechanism based on PCM mode.} Existing Tile-level ROI encryption methods can effectively prevent encryption diffusion by using the independent parallelism technology inherent in H.265/HEVC. However, the mutual reference of predictions in CU-level encryption leads to severe diffusion problems. By judiciously incorporating the PCM mode, we successfully suppressed the propagation of encryption-induced distortions into non-ROI regions, thereby ensuring that the background remains visually intact.

\item \textbf{Experimental comparison of different encryption strategies.} Extensive experimental results demonstrate that, compared with traditional Tile-level ROI encryption schemes, the proposed method achieves high-precision protection of ROI regions while effectively preventing the propagation of encryption-induced perturbations. Moreover, the proposed scheme introduces strong localized perturbations within the ROI, thereby providing robust and reliable protection of sensitive regions in the video.
\end{itemize}

The rest of this paper is organized as follows: Section \ref{sec2} reviews the related work. Section \ref{sec3} discusses the preliminaries. Section \ref{sec4} presents the proposed scheme. Section \ref{sec5} provides the experimental results and analysis. Finally, Section \ref{sec6} draws the conclusions.

\section{Related Work}\label{sec2}
\subsection{ROI Naive Encryption}
In the field of video privacy protection, early studies predominantly relied on naive encryption approaches, directly scrambling pixels or flipping syntax-compliant symbols within the ROI. For example, Dufaux \textit{et al}.\cite{dufaux2008scrambling} proposed two types of naive encryption methods for the MPEG-4 video surveillance scenario. One is transform coefficient sign flipping, which pseudo-randomly flips the signs of Alternating Current (AC)  coefficients of $4\times4$ blocks in the ROI. The other is bitstream bit flipping, which directly parses the MPEG-4 bitstream and pseudo-randomly flips the last bit of the Variable-Length Coding (VLC) codeword corresponding to AC coefficients. This study verified the practicability of ROI encryption in surveillance systems, but failed to solve the drift problem caused by inter-frame prediction. To address this issue, Dufaux \textit{et al}.\cite{dufaux2008h} utilized the Flexible Macroblock Ordering (FMO) mechanism in the H.264/AVC standard, dividing the ROI and background into independent slice groups and forcing background macroblocks to adopt intra coding. This completely eliminates prediction drift, enabling unauthorized users to clearly view non-sensitive regions. Carrillo \textit{et al}. \cite{carrillo2008compression} proposed a compression-independent block permutation encryption scheme, which performs pixel-level block scrambling on the detected ROI using pseudo-random permutation. Unlike video standard-based schemes, this method does not rely on specific compression algorithms and can maintain encryption effects during video transcoding and re-encoding processes, thus having good cross-platform applicability. 

Subsequently, Unterweger \textit{et al}. \cite{unterweger2016building} proposed a post-compression ROI encryption framework. This method directly applies Least Significant Bit (LSB) encryption to the differences in direct current coefficients of the ROI region in the bitstream after compression, avoiding the overhead caused by transcoding, but there is a risk of encryption leakage. To meet the lightweight requirement, Zhang \textit{et al}. \cite{zhang2018lightweight} proposed a lightweight ROI encryption scheme based on layered cellular automata (LCA). This scheme organizes the ROI extracted into binary blocks and implements parallel encryption using the reversible rules and shift transformation of LCA. This method can effectively resist known-plaintext attacks and differential attacks, and has good security while ensuring real-time performance. In recent years, ROI encryption research has been combined with deep learning, Hosny \textit{et al}. \cite{hosny2022privacy} combined Deep Convolutional Neural Network (DCNN) detection with block scrambling encryption. Their method uses YOLOv3 to detect and expand facial regions, and realizes ROI encryption through fast block scrambling and a key generation mechanism based on chaotic Logistic mapping. This scheme supports independent encryption of multiple ROIs, further improving the accuracy and security of ROI localization. However, naive encryption is incompatible with video encoding, making it difficult to apply in real-world scenarios. Cho \textit{et al}. \cite{cho2024practical} adopted CNN face detector to locate the ROI region, and generated entity-specific keys combining the master key and ID to encrypt the ROI. To solve the missed detection in ROI recognition, they proposed predicting the ROI of missed frames based on the ROI positions of adjacent frames and performing supplementary encryption. This scheme can achieve accurate matching between entities and IDs within the threshold range of 200–240, and can cover privacy protection for 64-frame missed detection scenarios, but it will increase a certain amount of computational overhead.

\subsection{ROI Selective Encryption}
Due to its compatibility with encoding technologies, ROI selective encryption has gradually attracted widespread attention. Farajallah \textit{et al}. \cite{farajallah2015roi} first proposed ROI encryption algorithm for H.265/HEVC encoding technology. They proposed two schemes, namely, bitstream-level encryption and selective encryption, and suppressed the diffusion phenomenon by limiting the reference range of motion vectors for non-ROI. This work ensures bitstream compatibility. Subsequently, Tew \textit{et al}. \cite{tew2016region} proposed three ROI encryption techniques. This scheme achieves format compatibility by manipulating the CABAC Bin string, providing a lightweight solution for low-complexity scenarios. Taha \textit{et al}. \cite{taha2018end} further implemented end-to-end real-time ROI encryption on the Kvazaar platforms. This scheme extends the encrypted syntax elements to luminance and chrominance IPM, and maintains the IPM scanning direction unchanged through cyclic shift operations, achieving a certain improvement in real-time performance. In addition, Yu \textit{et al}. \cite{yu2023coding} used YOLO to accurately locate the ROI region, then they adopts $64\times64$ CTU granularity for encryption, lacking fine-grained coding unit control, and essentially fails to address the issue of ROI precision.

\section{Preliminaries}\label{sec3} 
\subsection{Object Detection}\label{subsec3A}
Object detection aims to simultaneously locate and recognize target instances from input images, and its core task can be formalized as joint prediction of target categories and spatial positions. The process typically consists of three key stages: feature extraction, candidate prediction, and result refinement. Let the input image be $I\in\mathbb{R}^{H\times W\times C}$, it is first processed by a deep convolutional neural network to extract multi-scale feature representations:
\begin{equation}
    \mathbf{F}=\Phi_{\mathrm{feat}}(I)
\end{equation}
where $\Phi_{\mathrm{feat}}(\cdot)$ denotes the backbone network combined with a feature pyramid structure. The resulting feature maps $\mathbf{F}$ encode rich semantic information at multiple spatial resolutions.

Afterwards, object prediction is performed in the feature space. The detection head directly operates on the feature maps $\mathbf{F}$ to regress object locations and predict class probabilities, producing a set of candidate detections:
\begin{equation}
    \mathcal{D}=\{(b_i,c_i,s_i)\}_{i=1}^N
\end{equation}
where $b_i=(x_i,y_i,w_i,h_i)$ represents the bounding box parameters of the  $i$-th candidate object,  $c_i$ denotes the corresponding object category, and $s_i$ is the confidence score, $N$ denotes the number of detected objects in the current image. This process can be compactly expressed as:
\begin{equation}
    \mathcal{D}=\Phi_{\mathrm{det}}(\mathbf{F})
\end{equation}
with $\Phi_\mathrm{det}(\cdot)$ representing the detection head.

Finally, redundant detections are removed through confidence thresholding and Non-Maximum Suppression (NMS), yielding the final set of detected objects:
\begin{equation}
    \mathcal{B}=\mathrm{NMS}(\mathcal{D},\tau)
\end{equation}
where $\tau$ denotes the overlap threshold, and $\mathcal{B}=\{b_k\}_{k=1}^K$ is the final set of object bounding boxes. where $K$ denotes the number of final detected objects after non-maximum suppression. By integrating the above steps, the object detection module can be uniformly abstracted as an object detection function:
\begin{equation}
    \mathcal{B}=F_{\det}(I)
\end{equation}
where $F_{\det}(\cdot)$ defines the mapping from an input image to the detected object set.

\subsection{Prompt Segmentation}\label{subsec3B}
Prompt segmentation aims to produce accurate pixel-level segmentation of target regions under the guidance of prior prompts. Unlike conventional semantic segmentation, prompt segmentation introduces spatial or semantic priors to direct the model’s attention to specific objects, thus improving segmentation accuracy and robustness. In prompt segmentation, a set of prompts is first constructed based on the detection results:

\begin{equation}
    \mathcal{P}=\Psi(\mathcal{B})
\end{equation}
where $\Psi(\mathcal{\cdot})$ denotes the prompt generation function that converts bounding box information into prompt representations compatible with the segmentation model. Then, the input image $I$ and the prompt set $\mathcal{P}$ are separately encoded:
\begin{equation}
    \mathbf{F}_I=\Phi_{\mathrm{img}}(I),\quad\mathbf{F}_P=\Phi_{\mathrm{prompt}}(\mathcal{P})
\end{equation}
where $\Phi_{\mathrm{img}}(\cdot)$ is the image encoder and $\Phi_{\mathrm{prompt}}(\mathcal{\cdot})$ is the prompt encoder. The two feature representations are typically aligned in a shared embedding space.

After encoding, the image features and prompt features are fused to guide the segmentation network toward the prompted regions:
\begin{equation}
\mathbf{F}_{\mathrm{fusion}}=\Gamma(\mathbf{F}_{I},\mathbf{F}_{P})
\end{equation}
where $\Gamma(\cdot)$ denotes the feature fusion operator. Based on the fused features, the segmentation decoder predicts pixel-wise segmentation masks:

\begin{equation}
    M=\Phi_{\mathrm{seg}}(\mathbf{F}_\mathrm{fusion})
\end{equation}
where $M\in\{0,1\}^{H\times W}$ represents a binary segmentation mask of the target region. For multi-object or multi-candidate scenarios, the predicted masks are further refined and filtered according to their consistency with the prompts or confidence scores, yielding the final segmentation mask set:
\begin{equation}
    \mathcal{M}=\Omega(M,\mathcal{P})
\end{equation}
where $\Omega(\cdot)$ denotes mask selection and refinement operations. By integrating the above steps, the prompt segmentation module can be uniformly abstracted as:
\begin{equation}
    \mathcal{M}=F_{\mathrm{seg}}(I,\mathcal{B})
\end{equation}
where $F_{\mathrm{seg}}(\cdot)$ defines the mapping from an input image and detection prompts to pixel-level segmentation masks.

\subsection{Analysis of Syntax Elements in H.265/HEVC}
Syntax elements serve as the fundamental data units of video bitstreams, which are decoded by reconstructing video frames through parsing these elements. H.265/HEVC adopts context based adaptive binary arithmetic coding (CABAC) to adapt to the numerical distribution and statistical characteristics of different elements. Four binarization methods are defined: Fixed-Length (FL) coding\cite{flynn2015overview}, Truncated Unary (TU) coding, $k$-order Exponential Golomb ($\mathrm{EG}_k$) coding, and $k$-ocrder Truncated Rice ($\mathrm{TR}_k$) coding. The resulting binary bitstream is categorized into regular mode and bypass mode based on statistical properties. Regular mode utilizes an adaptive probability model for encoding, while bypass mode employs a fixed equal-probability model\cite{marpe2010entropy}. Encryption typically does not increase bitrate. Combining CABAC's binarization and encoding modes, primary syntax elements can be classified into prediction, residual, and filtering categories, as shown in TABLE \ref{tab1}.
\begin{table}[htbp]
\centering
\caption{\small Main syntax elements of H.265/HEVC}
\begin{tabular}{cccc}
\hline
\hline
\textbf{Syntax element} & \textbf{Coding mode} & \textbf{binarization} & \textbf{category} \\ \midrule
Luma IPM &Regular &  $\mathrm{TR}_k$ & {Prediction} \\
Chroma IPM &Regular & $\mathrm{TR}_k$ & {Prediction}  \\
Merge index & Regular, Bypass  & FL & {Prediction}  \\
MVD sign & Bypass & FL & {Prediction}  \\
MVD value & Bypass & $\mathrm{EG}_k$ & {Prediction}  \\
MVPIdx & Regular &  FL & {Prediction}  \\
RefFrmIdx & Regular, Bypass & $\mathrm{EG}_k$ & {Prediction}  \\
\cmidrule{1-4}
Residual sign &Bypass &  FL & {Residual} \\
Residual value &Bypass &  $\mathrm{TR}_k$ & {Residual} \\
Delta QP value & Regular, Bypass & $\mathrm{EG}_k$ & {Residual} \\
\cmidrule{1-4}
SAO parameter &Bypass  & $\mathrm{EG}_k$ & Filtering \\
\hline
\hline
\end{tabular}
\label{tab1}
\end{table}

As seen from TABLE \ref{tab1}, the syntax elements MVD sign and value, Residual sign and value, and SAO parameters, which encoded in bypass mode form the basis of zero bit rate incremental encryption, while the other syntax elements in mixed mode and regular mode provide the possibility for implementing adjustable encryption strategies. Combined with the CABAC mechanism for multi syntax element joint encryption, it can ensure format compatibility and compression efficiency, and achieve adjustable security protection in different scenarios. The syntax elements selected for encryption in this article are: IPM, MVPIdx, MergeIdx, RefFrmIdx, the values
and signs of MVD, as well as residual coefficients. These syntax elements directly determine the prediction mode, motion vector direction and amplitude, reference frame selection, and transform domain residual reconstruction results. They have the characteristics of high sensitivity and low bit occupancy, and can maintain compatibility with HEVC format while ensuring security.

\subsection{Analysis of the Encoding Process and Diffusion Principle of H.265/HEVC}\label{subsec3D}
In H.265/HEVC, each CU undergoes prediction, transform, quantization, and entropy coding operations to generate the final compressed bitstream. Concurrently, for subsequent prediction steps, it is subjected to inverse quantization, inverse transform, and loop-filtering to reconstruct pixels prior to entropy coding. Because these reconstructed pixels serve as references for later coding units, any encryption perturbations introduced during the intra prediction and inter prediction stages will propagate beyond the ROI boundaries, ultimately causing  encryption diffusion into non-ROI. In the following, we analyze the encryption diffusion principle caused by H.265/HEVC prediction. Assuming the original predicted values of non-ROI CU is:
\begin{equation}
    X={PR}(R)
\end{equation}
where $R$ are the reference pixels from the ROI CUs, \({PR}(\cdot)\) is the prediction function, and the residual value $S$ is calculated by subtracting the predicted value $X$ from the original pixels $O$ as $S=O-X$. Then, the transform, quantization, inverse quantization, and inverse transform are performed to $S$ to obtain the reconstructed residual values $S^{\prime}$. Finally, the reconstructed pixels of non-ROI CUs $O^{\prime}$ are computed as:
\begin{equation}
O^{\prime}=X+S^{\prime}.
\end{equation}

However, If the reference pixels in the ROI have been encrypted, the predicted values will be changed as:
\begin{equation}
    X_e={PR}(R+\Delta)
\end{equation}
where \(\Delta\) is the obvious encryption perturbation of ROI pixels, and the reconstructed pixels after encryption of the non-ROI CUs $O^{\prime}_e$ are:
\begin{equation}
O^{\prime}_e=X_e+S^{\prime}.
\end{equation}

Therefore, the deviation of the reconstructed pixels after encryption  can be expressed as:
\begin{equation}
    \delta O=O^{\prime}_e-O^{\prime}={PR}(R+\Delta)-{PR}(R)
\end{equation}

In fact, during both intra and inter prediction, non-ROI regions adjacent to the ROI boundary may reference pixel values from the ROI. As a result, the encryption in the prediction process is highly likely to propagate to neighboring regions. Moreover, because prediction is performed in a recursive manner, this propagation effect can accumulate and progressively amplify over time. In summary, the intra prediction and intre prediction process collectively contribute to distortion diffusion following CU-level encryption. These effects make high-precision ROI encryption a significant challenge in video privacy protection.

\section{Proposed Scheme}\label{sec4}
\begin{figure}[h]
    \centering
    \includegraphics[width=1\linewidth]{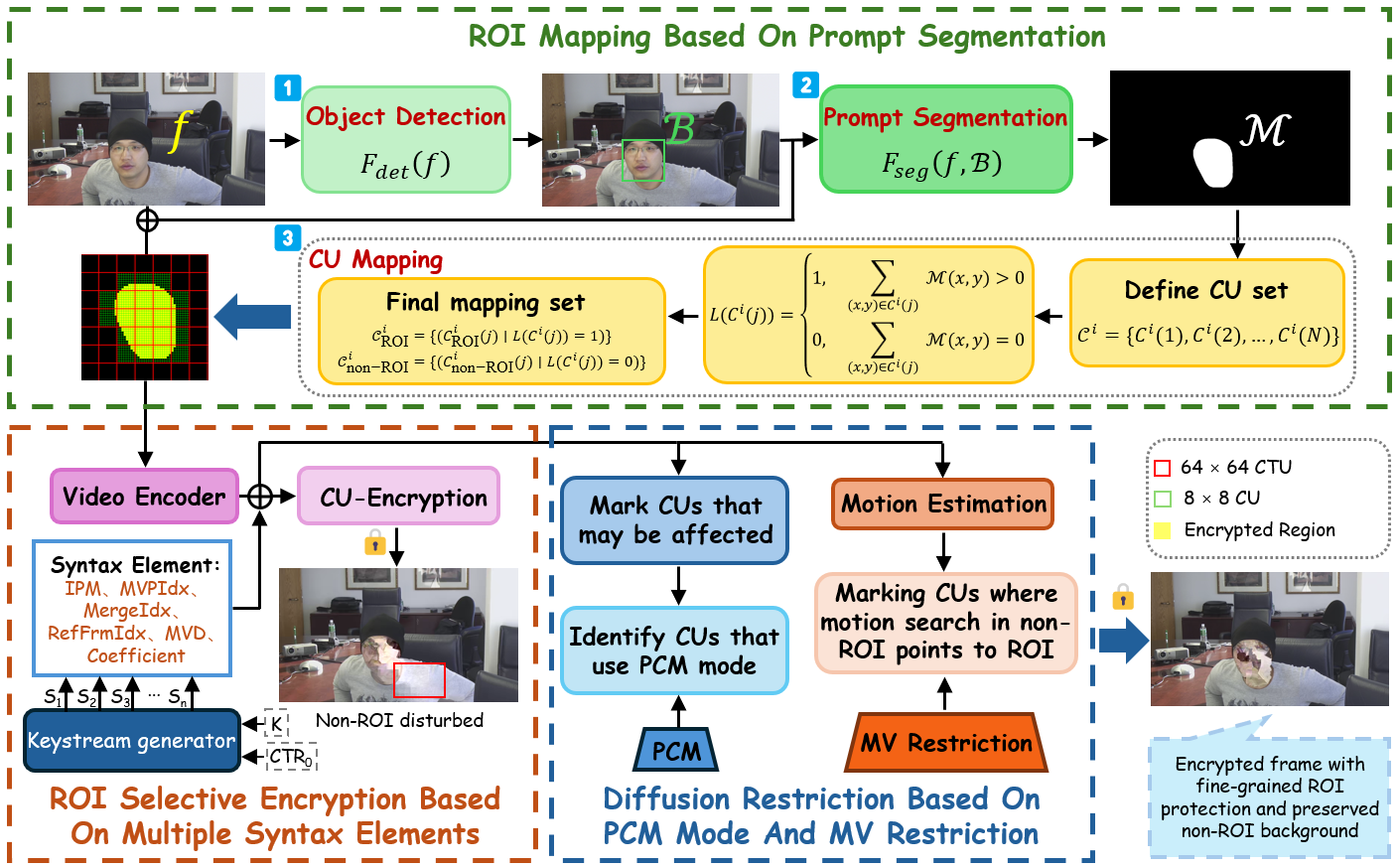}
    \caption{The proposed encryption scheme.}
    \label{fig:framework}
\end{figure}

The overall framework of the proposed scheme is shown in Figure \ref{fig:framework}. The proposed scheme is divided into three modules: ROI mapping based on prompt segmentation, ROI selective encryption based on multiple syntax elements, and diffusion isolation based on PCM mode and MV restriction. Among them, the ROI mapping consists of object detection, prompt segmentation and CU mapping. The ROI masks generated by this module serve as input for the ROI selective encryption module. Then, to address encryption diffusion issues, the diffusion isolation module produces the final encrypted results. The specific implementation steps are described below.

\subsection{ROI Mapping Based on Prompt Segmentation}\label{sec3a}

Our proposed ROI mapping framework consists of three parts: object detection, prompt segmentation, and CU mapping. 
\subsubsection{Object Detection}
Firstly, for the current input video frame $f$, Through the object detection function $F_{det}(\cdot)$, it predicts and locates the regions of interest within the video frame, and outputs the set of candidate object boxes for the current frame $\mathcal{B}$:
\begin{equation}
    \mathcal{B}=F_{det}(f)
\end{equation}

\subsubsection{Prompt Segmentation}

After obtaining the target bounding box $\mathcal{B}$ of the video frame, traditional rectangular boxes cannot accurately describe the edges of complex targets. Therefore, we use the target bounding box $\mathcal{B}$ as a spatial prior prompt to guide the segmentation model to perform precise pixel-level segmentation on the current video frame. Through the prompt segmentation function $F_{seg}(\cdot)$, a pixel-level mask corresponding to the target region of the current video frame is generated:
\begin{equation}
    \mathcal{M}=F_{seg}(f,\mathcal{B})
\end{equation}

\subsubsection{CU Mapping}

After completing pixel-level ROI detection, we propose a CU mapping strategy to  map the mask information to the H.265/HEVC coding structure. Specifically, H.265/HEVC encoder takes the CTU as the basic starting point and recursively divides it into CUs through a quadtree structure. Let the CU set of frame $i$ be:
\begin{equation}
\mathcal{C}^i=\{C^i(1),C^i(2),\ldots,C^i(N)\}
\end{equation}
where $C^i(j)$ represents that the $j$-th CU in frame $i$ with the size of $w_j \times h_j$, and the pixel coordinates in the top-left corner of $C^i(j)$ are ($x_j$, $y_j$). Traverse all CUs from frame $i$, if there exists at least one pixel $(x,y)\in C^i(j)$ such that $\mathcal{M}(x,y)=1$, the CU is labeled as an ROI-CU; otherwise, it is labeled as a non-ROI CU. Accordingly, the CU-level ROI label $L(C^i(j))\in\{0,1\}$ is defined as:
\begin{equation}
L(C^i(j))=
\begin{cases}
1, & \sum_{(x,y)\in C^i(j)}\mathcal{M}(x,y)>0 \\
0, & \sum_{(x,y)\in C^i(j)}\mathcal{M}(x,y)=0
\end{cases}
\end{equation}
Where, $L(C^i(j))=1$ indicates that $C^i(j)$ belongs to the ROI, while $L(C^i(j))=0$ indicates that $C^i(j)$ belongs to the non-ROI. The final mapping ROI CU set and non-ROI CU set result of frame $i$ is represented as:
\begin{equation}
\begin{aligned}
    \mathcal{C}_{\text{ROI}}^i=\{(C_{\text{ROI}}^i(j) \mid L(C^i(j))=1)\} \\
\mathcal{C}_{\text{non-ROI}}^i=\{(C_{\text{non-ROI}}^i(j) \mid L(C^i(j))=0)\}
\end{aligned}
\end{equation}

\subsection{ROI Selective Encryption Based on Multiple Syntax Elements}

After ROI mapping, assume the current frame is the $i$-th frame, selective encryption is performed on IPM, MVPIdx, MergeIdx, RefFrmIdx, MVD, and coefficients of ROI CU set $\mathcal{C}_{\text{ROI}}^i$. To balance security and bitrate overhead, AES-CTR is employed to generate the pseudo-random keystream \cite{lipmaa2000ctr}. With key $K$ and initial counter \(CTR_0\), the counter and keystream are given by:
\begin{equation}
    CTR_{n+1}=CTR_n+1
\end{equation}
\begin{equation}
    S_n=AES(K,CTR_n)
\end{equation}
where \(S_n\) is the keystream fragment for the $n$-th data block, subsequent encryption of each syntax element is as follows.

\subsubsection{Intra Prediction Mode Encryption}
The intra prediction mode (IPM) includes Luma IPM and Chroma IPM. If the Luma IPM falls within this candidate list, the corresponding mode index is encrypted as follows:
\begin{equation}
    enc\_preIdx_{c_{\text{ROI}}^i}=(preIdx_{c_{\text{ROI}}^i}+S_t)\%3, \quad 0 \leq \mathit{S_t} \leq 3.
\end{equation}

If Luma IPM is not in the candidate list, perform XOR operation on the 5-bit offset as:
\begin{equation}
    enc\_dir_{c_{\text{ROI}}^i}=dir_{c_{\text{ROI}}^i}\oplus S_t, \quad 0 \leq \mathit{S_t} \leq 31.
\end{equation}

Whereas chroma IPM includes five prediction modes, perform XOR operation directly on the mode number during encryption:
\begin{equation}
\begin{aligned}
    & enc\_uiDirChroma_{c_{\text{ROI}}^i}= \\ & (enc\_uiDirChroma_{c_{\text{ROI}}^i}+S_t)\%3, 
    \quad 0 \leq \mathit{S_t} \leq 3.
\end{aligned}
\end{equation}

\subsubsection{Motion Vector Prediction Index Encryption}
When using Advanced Motion Vector Prediction (AMVP), the selected MVP index is encrypted as:
\begin{equation}
    enc\_MVPIdx_{c_{\text{ROI}}^i}=MVPIdx_{c_{\text{ROI}}^i}\oplus S_t, \quad 0 \leq S_t \leq 1.
\end{equation}

\subsubsection{Merge Index Encryption}
In Merge mode, the encoder collects a set of candidate predicted motion information from surrounding blocks, the index associated with this list is encrypted as:
\begin{equation}
    enc\_MergeIdx_{c_{\text{ROI}}^i}=(MergeIdx_{c_{\text{ROI}}^i}+S_t)\%5,\,\,\,\,\! 0 \leq S_t \leq 3.
\end{equation}

\subsubsection{Reference Frame Index Encryption}
The value range of the reference frame index (RefFrmIdx) is determined by the number of frames $RN$ currently stored in the reference frame buffer. The encryption of RefFrmIdx is manifested as:
\begin{equation}
    enc\_RFI_{c_{\text{ROI}}^i}=
\begin{cases}
RFI_{c_{\text{ROI}}^i}\oplus S_t &  0\leq S_t\leq1  \, \\ &if\,\,\  RN=2\\
(RFI_{c_{\text{ROI}}^i}+S_t)\%3 & 0\leq S_t\leq1  \,\\ &if\,\,\  RN=3\\
RFI_{c_{\text{ROI}}^i}\oplus S_t &  0\leq S_t\leq3 \,\\ &if\,\,\  RN=4.
\end{cases}
\end{equation}

\subsubsection{Motion Vector Difference Encryption}
The Motion Vector Difference (MVD) includes both horizontal and vertical components. We encrypt the symbols of the horizontal component and the vertical component as follows:
\begin{equation}
\begin{aligned}
   enc\_Horsign_{c_{\text{ROI}}^i}=&Horsign_{c_{\text{ROI}}^i}\oplus S_t,\,\,\,\,\! 0 \leq S_t \leq 1,\\
   enc\_Versign_{c_{\text{ROI}}^i}=&Versign_{c_{\text{ROI}}^i}\oplus S_t,\,\,\,\,\! 0 \leq S_t \leq 1.
\end{aligned}
\end{equation}

\subsubsection{Residual Coefficient Sign and Value Encryption}
The residual coefficient consists of sign and value. The sign is encrypted as:
\begin{equation}
    enc\_coefsign_{c_{\text{ROI}}^i}=coefsign_{c_{\text{ROI}}^i}\oplus S_t,\,\,\,\,\! 0 \leq S_t \leq 1.
\end{equation}

The value consists of $CoefBaseLevel$ and $CoefRemainingLevel$. we only encrypts the suffix of the $CoefRemainingLevel$ as:
\begin{equation}
    enc\_coefsuffix_{c_{\text{ROI}}^i}=coefsuffix_{c_{\text{ROI}}^i}\oplus S_t,\,\,\,\,\! 0 \leq S_t \leq 1.
\end{equation}

\subsection{Diffusion Isolation Based on PCM Mode and MV Restriction}
As mentioned in section \ref{subsec3D}, propagation driven by prediction mechanisms typically exhibits distinct spatial structures and directional characteristics in the reconstructed frames after decoding, enabling direct observation of the propagation regions. Therefore, this section will present solutions to the distortion propagation problem based on theoretical foundations derived from the intra prediction and inter prediction mechanism level.

\subsubsection{Intra prediction diffusion isolation based on PCM Mode}
\begin{figure}[h]
    \centering
    \includegraphics[width=0.65\linewidth]{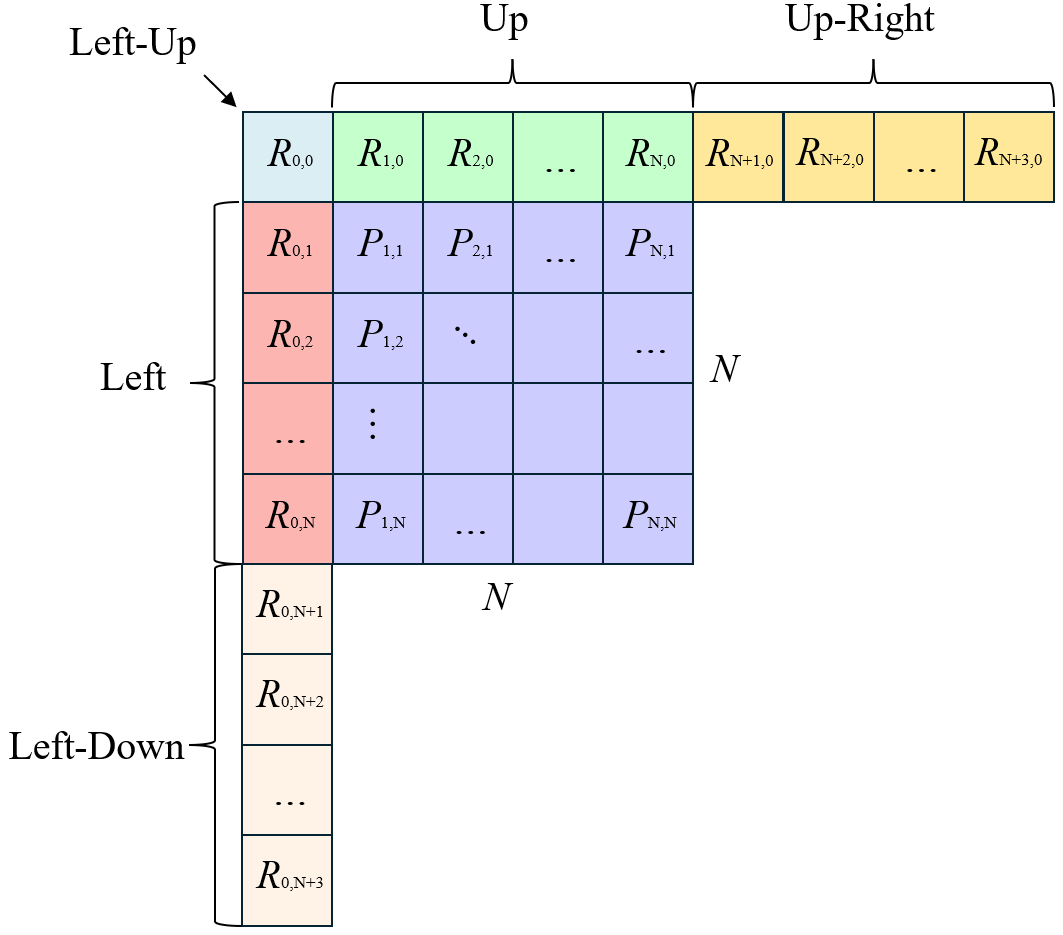}
    \caption{\small Reference pixel position for intra prediction}
    \label{fig3}
\end{figure}

In the intra prediction process, the reconstructed pixel values of CU are highly dependent on the prediction results of adjacent reconstructed blocks \cite{6316136}. The predicted pixels are usually obtained through a reference pixel set, which includes reconstructed samples in five directions: Up, Up-Right, Left, Left-up, and Left-Down, as shown in Figure \ref{fig3}. Therefore, when performing intra prediction, the encoder will use these neighboring pixels for linear interpolation or angle prediction, thereby generating the prediction block for the current CU. This neighborhood based dependency is an important mechanism for H.265/HEVC to improve compression rate, but when the pixels within the ROI are perturbed after encryption, these encrypted pixels may propagate into the non-ROI through the aforementioned five directions.

\begin{figure}[h]
    \centering
    \includegraphics[width=0.7\linewidth]{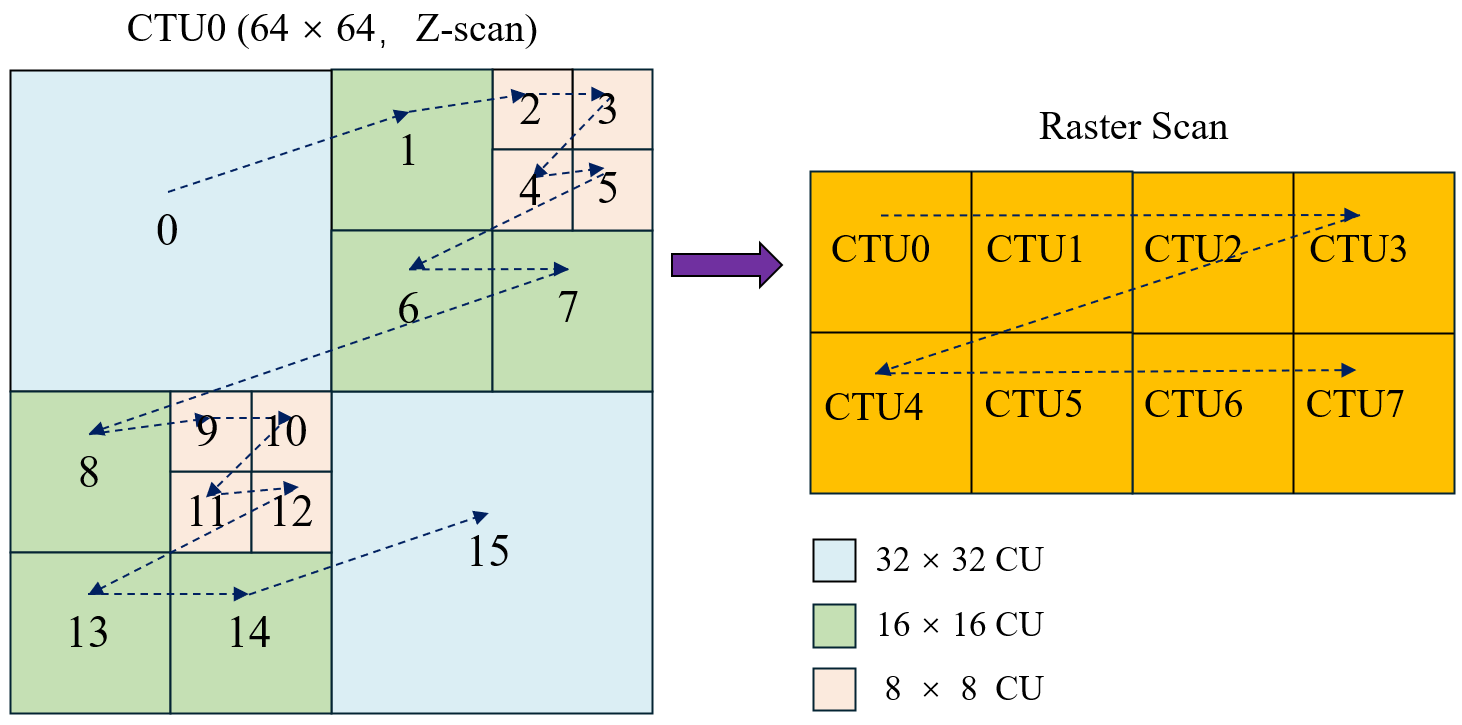}
    \caption{\small CU scanning order in H.265/HEVC}
    \label{fig:scan}
\end{figure}

To address this issue, we proposes a intra prediction diffusion isolation mechanism based on Mode Pulse Code Modulation (PCM). PCM is a special encoding mode in the H.265/HEVC standard \cite{Rao2014HighEV}. When the CU adopts PCM mode, the encoder completely bypasses the prediction, transformation, and quantization processes, and directly performs lossless encoding on the original pixel values \cite{Sze2014HighEV}. The specific diffusion isolation method is as follows:

We first obtain the ROI CU set in the $i$-th frame $\mathcal{C}_{\text{ROI}}^i$ by mapping rules of Section \ref{sec3a}. H.265/HEVC standard processes each CTU in a raster order from left to right and top to bottom. Within each CTU, the CU is traversed in Z-scan order based on the quadtree partition structure, as shown in Figure \ref{fig:scan}. For example, suppose CU number 11 belongs to the ROI. When it is encoded, CUs numbered 0-10 have already been encoded sequentially. Even if CU number 11 is disturbed, it will not affect the CUs above and to its left. However, CUs numbered 12, 13, and 14 as the neighboring CUs to the right and below CU number 11 may use the reconstruction result of CU number 11 as a reference during subsequent encoding.

In summary, under this scanning strategy, when encoding the CU above or to the left of the ROI, the ROI has not yet been encoded and therefore will not be affected. However, After the ROI is encoded and encrypted, the neighboring CUs in the right, lower, and diagonal directions may reference the reconstructed pixels of ROI during the prediction process according to the reference mechanism of H.265/HEVC, resulting in encryption diffusion phenomenon. From this, it can be seen that the diffusion path mainly occurs in the five directions of the ROI: Right, Down, Right-Down, Right-Up, and Left-Down. Based on the above analysis, we define the potential reference neighborhood CUs for $C_{\text{ROI}}^i(j)$ as the five locations listed in Table \ref{tabCr}.

\begin{table}[htbp]
\centering
\footnotesize
\setlength{\tabcolsep}{4pt} 
\caption{\small Neighborhood regions of ROI potentially affected by ROI}\label{tabCr}
\begin{tabular}{ccc}  
\hline
\hline
Location   & Pixel coordinate in the top-left corner & Size    \\ \hline
$C_{\text{ROI}}^i(j)\_{\text{Right}}$     & $(x_j+w_j,y_j)$    & $w_j\times h_j$    \\
$C_{\text{ROI}}^i(j)\_\text{Down}$     & $(x_j,y_j+h_j)$    & $w_j\times h_j$    \\
$C_{\text{ROI}}^i(j)\_\text{Right-Down}$  & $(x_j+w_j,y_j+h_j)$  & $w_j\times h_j$ \\
$C_{\text{ROI}}^i(j)\_\text{Right-Up}$    & $(x_j+w_j,y_j-h_j)$  &  $w_j\times h_j$ \\
$C_{\text{ROI}}^i(j)\_\text{Left-Down}$  & $(x_j-w_j,y_j+h_j)$  &  $w_j\times h_j$ \\ \hline
\hline
\end{tabular}%
\end{table}

Define the set of five CUs in Table \ref{tabCr} as $\mathcal{C}_{\text{ROI}}^i(j)\_\text{Nb}$. If there exists a non-ROI CU $C_{\text{non-ROI}}^i(k)$ that satisfies $C_{\text{non-ROI}}^i(k) \in \mathcal{C}_{\text{ROI}}^i(j)\_\text{Nb}$. It is considered that the $C_{\text{non-ROI}}^i(k)$ is located within the ROI prediction reference neighborhood and has potential encryption diffusion risk. In order to control the bit rate overhead while ensuring diffusion isolation, we only forces CUs that meet the following conditions to use PCM encoding mode. It can be expressed as:
\begin{equation}
    PCM(C_{\text{non-ROI}}^i(k))=
\begin{cases}
1, &  C_{\text{non-ROI}}^i(k) \in \mathcal{C}_{\text{ROI}}^i(j)\_\text{Nb}, \\& w_k=h_k=8 \\
0, & \mathrm{otherwise,} 
\end{cases}
\end{equation}
where \(\text{PCM}(C_{\text{non-ROI}}^i(k))=1\) indicates that the CU is marked as a forced PCM encoding mode, \(w_k=h_k=8\) indicates that the size of CU is fixed at $8 \times 8$ to meet the minimum PCM block size specified by the standard. The final set of mandatory PCM blocks is:
\begin{equation}
   \mathcal{C}_{\text{PCM}}=\{C_{\text{non-ROI}}^i(k)\ |\ \text{PCM}(C_{\text{non-ROI}}^i(k))=1\}
\end{equation}

All CUs in the set \(\mathcal{C}_{\text{PCM}}\) will skip the conventional rate distortion optimization and mode decision process during the encoding process and directly enter the PCM encoding process, thereby cutting off the diffusion effect caused by encryption.

\subsubsection{Inter prediction diffusion isolation based on MV restriction}
Inter prediction utilizes the temporal correlation of videos and employs motion compensation techniques to find the best matching reference block for current block from previously encoded frame. In the inter frame prediction process, once the non-ROI CU references the reconstructed CUs in ROI through the motion estimation process, the perturbations introduced by encryption in the ROI will propagate along the prediction link in the time dimension to the non-ROI, forming a cross region encryption diffusion phenomenon.

To address this issue, we further introduce an inter prediction diffusion isolation mechanism based on MV restriction. For a non-ROI CU $C_{\text{non-ROI}}^{i}(m)$ in frame $i$, let its motion vector set be:
\begin{equation}
\mathbf{v}^i(m)=(\Delta x(m),\Delta y(m)).
\end{equation}

When $C_{\text{non-ROI}}^{i}(m)$ is mapped to the reference frame $r$ through motion vector set $\mathbf{v}^i(m)$, its reference CU set in the reference frame $r$ is $\mathcal{C}^{r}(m)\_{\text{ref}}$. If $\mathcal{C}^{r}(m)\_{\text{ref}}$ intersects with the ROI CU set in the $r$-th frame $\mathcal{C}_{\text{ROI}}^r$, that is:
\begin{equation}
\mathcal{C}^{r}(m)\_{\text{ref}}\cap \mathcal{C}_{\text{ROI}}^{r}\neq \varnothing,
\end{equation}
the motion vector is considered to pose a risk of encryption diffusion. Therefore, this motion vector cannot be used for inter prediction; otherwise, it could be used for inter prediction. Accordingly, we define the validity of a motion vector as:
\begin{equation}
\Phi\!\left(C_{\text{non-ROI}}^{i}(m),\mathbf{v}^i(m)\right)=
\begin{cases}
1, & \mathcal{C}^{r}(m)\_{\text{ref}}\cap \mathcal{C}_{\text{ROI}}^{r}=\varnothing,\\
0, & \text{otherwise}.
\end{cases}
\end{equation}

Only valid motion vectors are retained for inter prediction, when $\Phi\!\left(C_{\text{non-ROI}}^{i}(m),\mathbf{v}^i(m)\right)=1$, non-ROI CU $C_{\text{non-ROI}}^{i}(m)$ are constrained to reference only non-ROI region in the reference frame $r$, thereby preventing encrypted distortions in the ROI from propagating to non-ROI along the temporal prediction path.

\section{Experiments}\label{sec5}
\subsection{Experimental Setup}
Our scheme is implemented in the H.265/HEVC reference software HM 16.9, with the computer configuration of Inter (R) Core (TM) i7-9750H CPU @2.60GHz, 16GB of memory, Windows 11 operating system, Microsoft Visual Studio 2010, MATLAB 2021a, and OpenCV 2.4.7. The configuration file used in the encoder is ``encoder\_lowdelay\_main", we set PCM mode to 1, Group of Pictures (GOP) is set to ``IBBB", QP is set to 24, respectively. The model for object detection is YOLOv8, the prompt segmentation model is SAM. We select five YUV sequences containing faces for testing, with resolutions ranging from $352\times288$ to $1920\times1080$, which are shown in TABLE \ref{tab2}. In order to comprehensively evaluate the performance of the proposed ROI selective encryption scheme in various aspects, we cite the ROI encryption evaluation benchmark proposed by Zhang et al. \cite{zhang2025visual} to evaluate the ROI encryption accuracy and ROI perturbation effect.

\begin{table}[htbp]
\centering
\footnotesize
\setlength{\tabcolsep}{18pt} 
\caption{\small Test video sequences used in experiments}\label{tab2}
\begin{tabular}{ccc}  
\hline
\hline
Sequence   & Resolution & FPS    \\ \hline
Akiyo       & $352\times288$    & 60    \\
PartyScene     & $832\times480$    & 60 \\
Johnny        & $1280\times720$   &  60  \\
Vidyo4        & $1280\times720$  &  60  \\
Kimono  &$1920\times1080$  &  60\\ \hline
\hline
\end{tabular}%
\end{table}

\begin{figure}[h]
    \centering
    \subfloat[Detection]{
        \includegraphics[width=0.42\columnwidth]{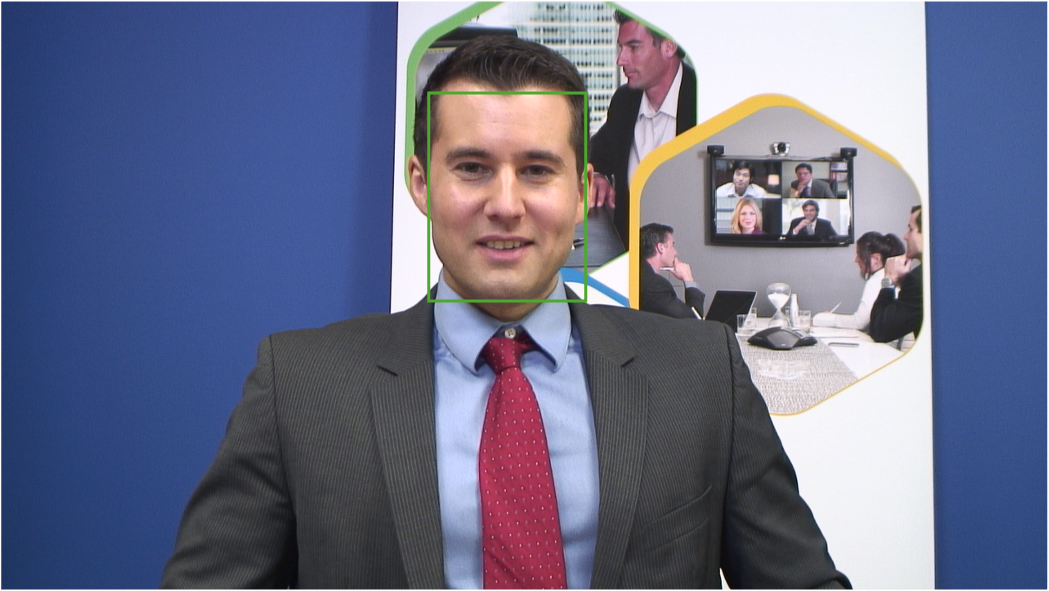}
        \label{fig:det_result}
    }
    \hfill
    \subfloat[Segmentation]{
        \includegraphics[width=0.42\columnwidth]{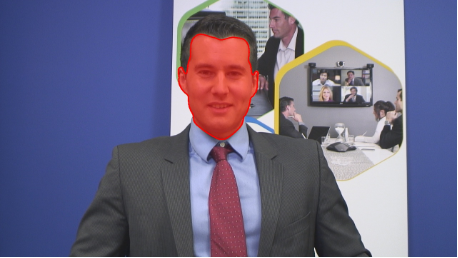}
        \label{fig:seg_result}
    }
    \caption{Results after detection and segmentation.}
    \label{fig:det_seg_result}
\end{figure}

In this experiment, the face is selected as the sensitive area for encryption. Firstly, the WiderFace dataset is used for face recognition training. In order to improve the accuracy of the model for face detection, this dataset includes various complex scene transformations. The trained model can accurately output rectangular boxes that strictly fit the face region, and then use this output as input prompt words for the segmentation model. It can obtain pixel level masks that highly match the face contour. The visualization of object detection is shown in Figure \ref{fig:det_seg_result}(1) and segmentation is shown in Figure \ref{fig:det_seg_result}(2).

\subsection{Experimental Results}
\begin{figure}[h]
    \centering
    \subfloat[Original]{
        \includegraphics[width=0.28\columnwidth]{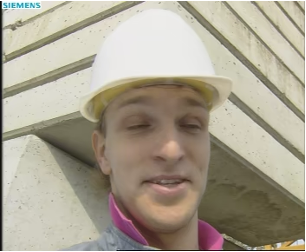}
        \label{fig:ori}
    }
    \hfill
    \subfloat[Encrypted]{
        \includegraphics[width=0.28\columnwidth]{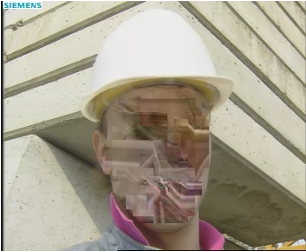}
        \label{fig:enc}
    }
    \hfill
    \subfloat[Decrypted]{
        \includegraphics[width=0.28\columnwidth]{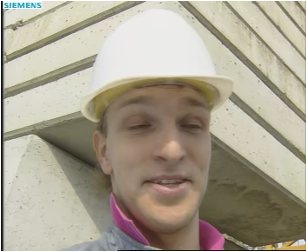}
        \label{fig:dec}
    }
    \caption{Visual examples of the proposed scheme.}
    \label{fig:roi_example}
\end{figure}

To visually demonstrate the effectiveness of the proposed ROI encryption scheme, Figure \ref{fig:roi_example} shows the encrypted and decrypted sample frames of the video sequence foreman under the condition of QP=24. Specifically, the first frames of the video sequence were selected as examples to visually present the encryption and decryption effects. The experimental results show that the video stream generated by the ROI encryption scheme proposed in this paper can be correctly parsed by the HM standard decoder, indicating that this scheme fully meets the format compatibility requirements of H.265/HEVC. Meanwhile, the encrypted sample frames exhibit obvious fine-grained distortion features, making it difficult to directly identify sensitive ROI content in the original video.

\subsection{Performance Analysis}
We compare three Full-frame encryption schemes proposed by Sheng et al. \cite{sheng2024chaos}, Fu et al. \cite{sheng2024contentt}, Tie et al. \cite{sheng2024contents} and the ROI encryption scheme proposed by Zhang et al. \cite{zhang2025visual}, and Taha et al. \cite{taha2018end}. Due to the unstable encryption performance of each frame and the different number of frames in each test sequence, in order to ensure the consistency of experimental data, the experiment uniformly uses 50 frames to test the performance for comparison. The specific comparative data is as follows.

\subsubsection{ROI encryption accuracy results}
In order to quantitatively evaluate the accuracy of the proposed ROI encryption scheme in spatial positioning, we employ encryption accuracy indicator \(IoU_{avg}\) introduced by Zhang et al. \cite{zhang2025visual}. This metric is used to measure the spatial consistency between the actual encrypted CU set in the encoder and the real ROI region marked by the segmentation model. we compare our scheme with two ROI encryption scheme, Taha et al. \cite{taha2018end}, and Zhang et al. \cite{zhang2025visual}. The comparison results of \(IoU_{avg}\) for five test sequences are shown in Table \ref{tab:IoU}. 
\begin{table}[htbp]
\centering
\footnotesize
\caption{Average IoU of the compared schemes}
\label{tab:erd_values}
\centering 
\setlength{\tabcolsep}{12pt}
\begin{tabular}{lccc} 
\hline
\hline
\multirow{2}{*}{Sequence} & \multicolumn{3}{c}{IoU (\%)} \\ 
\cline{2-4} 
 & Taha~\cite{taha2018end} & Zhang~\cite{zhang2025visual} & Ours \\ 
\hline
Akiyo & 0.760 &\underline{0.845} &  \textbf{0.882}  \\
PartyScene & 0.770 &\underline{0.842} &  \textbf{0.879}  \\
Johnny & 0.785 &\underline{0.848} &  \textbf{0.895}  \\
Vidyo4 & 0.800 &\underline{0.850} &  \textbf{0.873}  \\
Kimono & 0.815 &\underline{0.852} &  \textbf{0.866}  \\ 
\hline
\hline
\end{tabular}
\label{tab:IoU}
\end{table}

Experimental results demonstrate that the proposed CU-level ROI encryption scheme outperforms both the $16\times16$ Tile-level encryption scheme proposed by Zhang et al. \cite{zhang2025visual} and the $32\times32$ Tile-level encryption scheme proposed by Taha et al. \cite{taha2018end} in terms of the \(IoU_{avg}\) metric. This is because Tile-level schemes are constrained by fixed block structures, making it difficult to align their encryption boundaries with the irregular semantic contours of the target. This often introduces over-encrypted regions, thereby limiting the upper bound of \(IoU_{avg}\). In contrast, our scheme utilizes the adaptively partitioned CUs from the H.265/HEVC encoding process as encryption units. Through precise CU mapping, it reduces the size of the boundary units to a fixed $8\times8$ dimension, enabling finer-grained encryption that better aligns with the target spatial structure. This mechanism effectively minimizes boundary errors, achieving high consistency between the encrypted region and the actual ROI.

To further evaluate the spatial granularity of different ROI encryption schemes, we additionally introduces Encryption Redundancy Rate ($ERR$) to measure the proportion of non-ROI that are unnecessarily included in the actual encrypted area, thereby reflecting how accurately the encrypted region fits the ground-truth ROI. Let $G$ denote the ground-truth ROI and $E$ denote the actual encrypted region. The encryption redundancy rate is defined as:
\begin{equation}
    ERR = \frac{\lvert E \rvert - \lvert E \cap G \rvert}{\lvert E \rvert}
\end{equation}
where $\lvert E \rvert$ represents the area of the actual encrypted region, and $ \lvert E \cap G \rvert$ denotes the overlapping area between the actual encrypted region and the ground-truth ROI. A lower $ERR$ indicates that the encrypted region is closer to the true ROI, with higher spatial precision. The comparison results are shown in Table \ref{tab:ERR}. The results show that the proposed scheme has a significantly lower $ERR$ than the other two comparative schemes, further verifying the superiority of this scheme in fine-grained ROI protection.

\begin{table}[htbp]
\centering
\footnotesize
\caption{Average ERR of the compared schemes}
\label{tab:erd_values}
\centering 
\setlength{\tabcolsep}{12pt}
\begin{tabular}{lccc} 
\hline
\hline
\multirow{2}{*}{Sequence} & \multicolumn{3}{c}{ERR (\%)} \\ 
\cline{2-4} 
 & Taha~\cite{taha2018end} & Zhang~\cite{zhang2025visual} & Ours \\ 
\hline
Akiyo & 0.745 &\underline{0.617} &  \textbf{0.191}  \\
PartyScene & 0.967 &\underline{0.870} &  \textbf{0.463}  \\
Johnny & 0.171 & 0.171 &  \textbf{0.125}  \\
Vidyo4 & 0.701 &\underline{0.462} &  \textbf{0.063}  \\
Kimono & 0.797 &\underline{0.636} &  \textbf{0.153}  \\ 
\hline
\hline
\end{tabular}
\label{tab:ERR}
\end{table}

\subsubsection{Visual comparative analysis of the effectiveness of diffusion isolation}
\begin{figure}[h]
\centering


\subfloat[\ Original]{
\includegraphics[width=0.28\columnwidth]{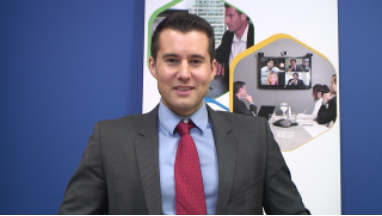}}
\hfill
\subfloat[\ No diffusion isolation]{
\includegraphics[width=0.28\columnwidth]{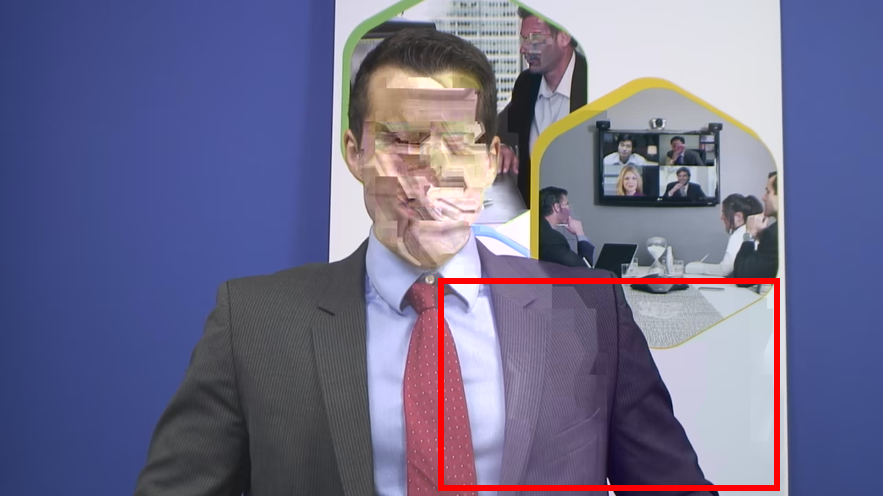}}
\hfill
\subfloat[\ Diffusion isolation]{
\includegraphics[width=0.28\columnwidth]{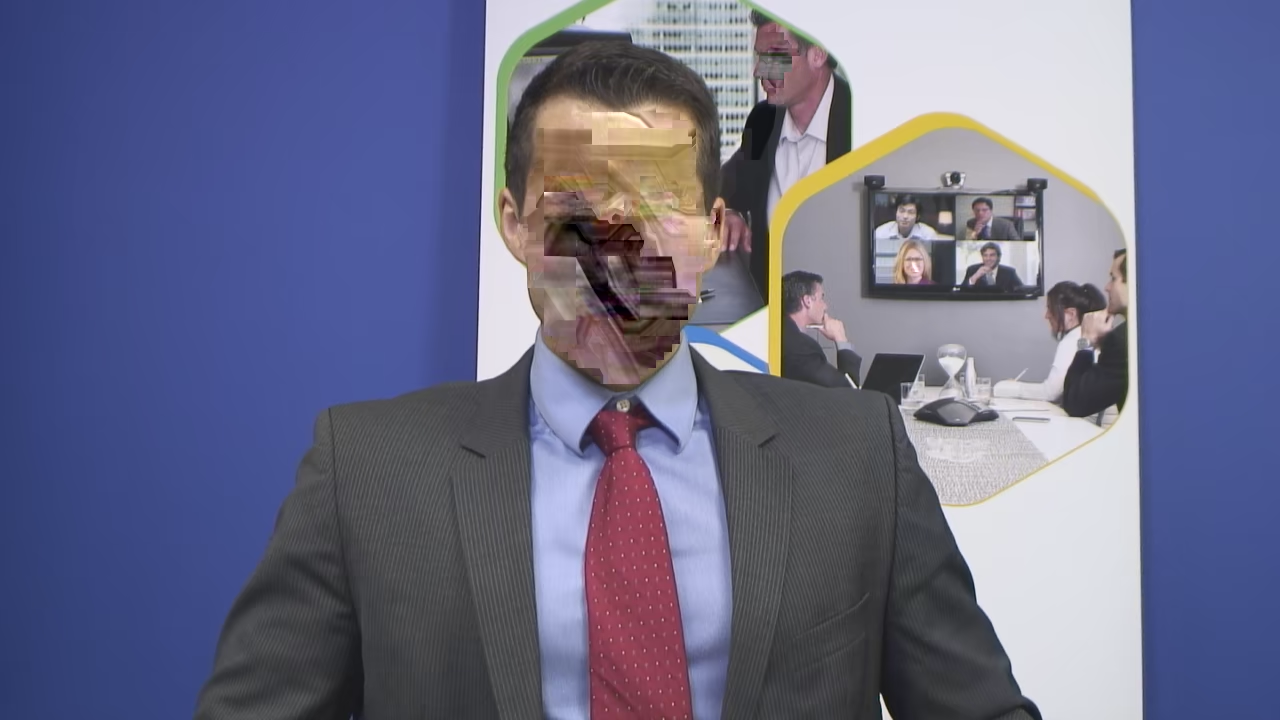}}

\subfloat[\ Original]{
\includegraphics[width=0.28\columnwidth]{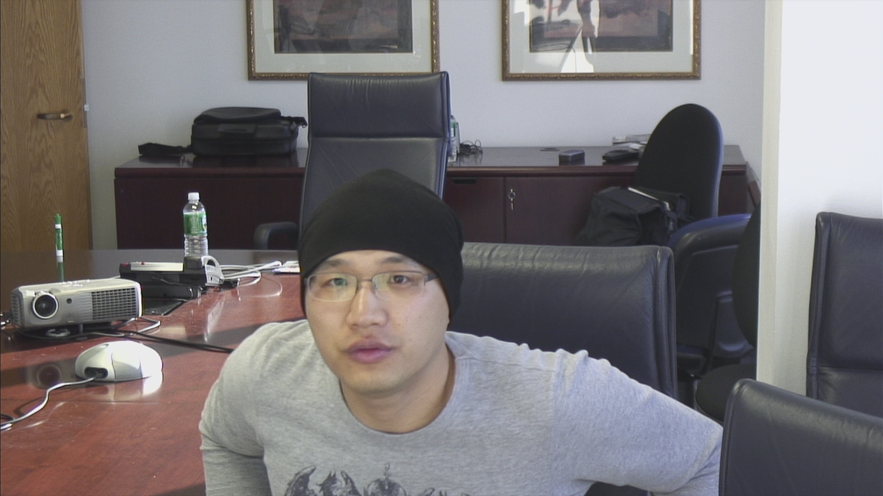}}
\hfill
\subfloat[\ No diffusion isolation]{
\includegraphics[width=0.28\columnwidth]{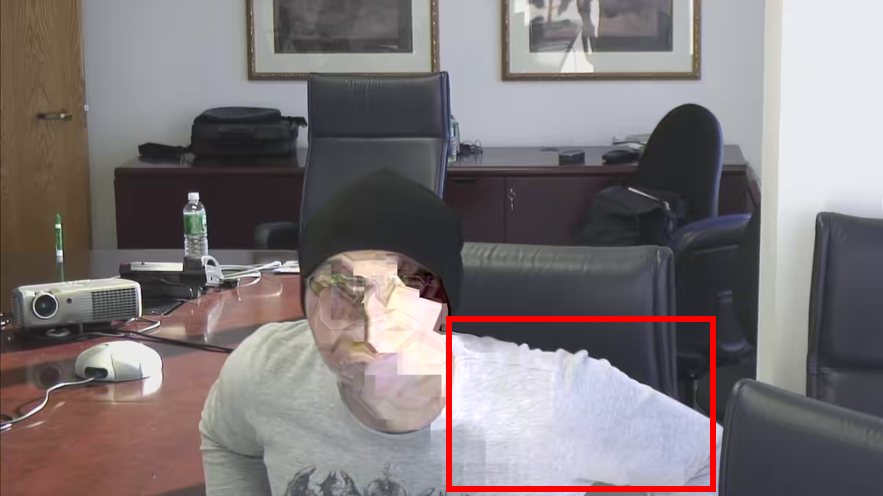}}
\hfill
\subfloat[\ Diffusion isolation]{
\includegraphics[width=0.28\columnwidth]{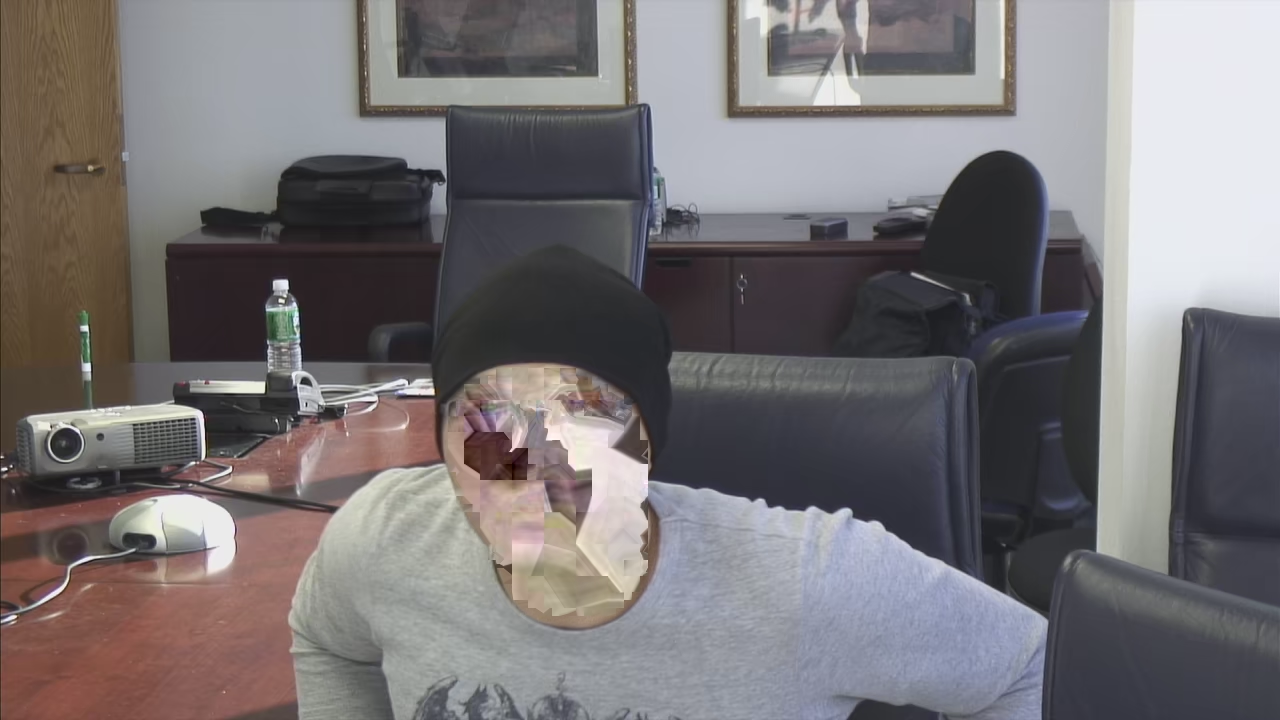}}

\caption{Visual ablation study of diffusion isolation. From left to right are the original frame, the encrypted frame without diffusion isolation, and the encrypted frame with diffusion isolation.}
\label{diff-comp}
\end{figure}

To visually verify the effectiveness of the proposed diffusion isolation mechanism, an ablation study is conducted by comparing the encrypted results without and with diffusion isolation. As shown in Figure \ref{diff-comp}, without diffusion isolation, the distortion introduced in the ROI propagates to surrounding non-ROI due to coding dependencies, resulting in noticeable visual degradation outside the ROI. In contrast, with the proposed diffusion isolation mechanism, the distortion is confined within the ROI, while the non-ROI remain intact. This demonstrates that the proposed mechanism can effectively suppress the spatial leakage of encryption distortion and improve the spatial reliability of CU-level ROI selective encryption.

\subsubsection{Comparison of ROI perturbation effects}
The comparison of ROI perturbation effects includes two main parts: subjective visual comparison and objective index comparison. The following is a detailed analysis of this part.

\begin{figure*}[h]
\centering
\captionsetup{font=footnotesize}
\captionsetup[subfloat]{font=footnotesize}

\subfloat[Frame \#1\\Original]{
\includegraphics[width=0.110\textwidth]{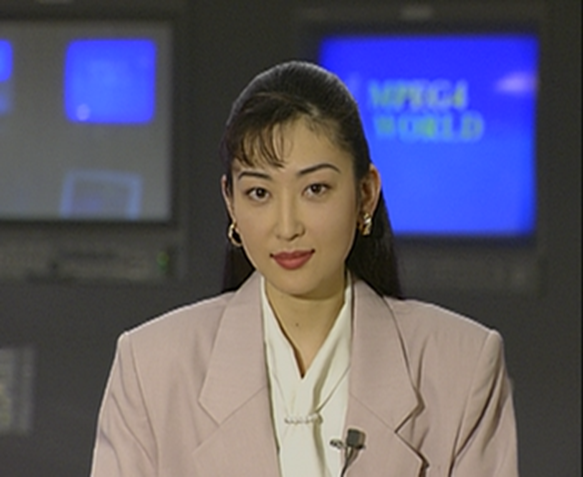}}
\hfill
\subfloat[Frame \#1\\Taha~\cite{taha2018end}]{
\includegraphics[width=0.110\textwidth]{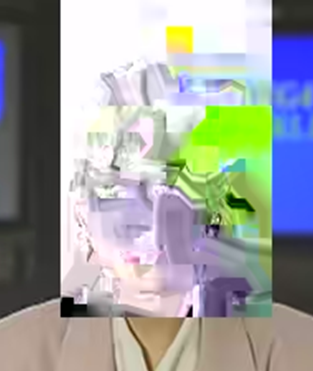}}
\hfill
\subfloat[Frame \#1\\Zhang~\cite{zhang2025visual}]{
\includegraphics[width=0.110\textwidth]{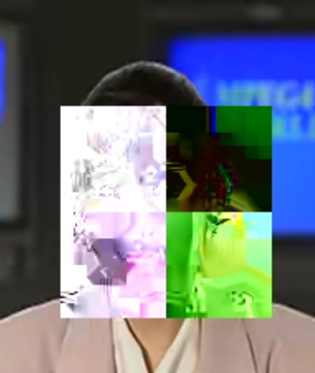}}
\hfill
\subfloat[Frame \#1\\Ours]{
\includegraphics[width=0.110\textwidth]{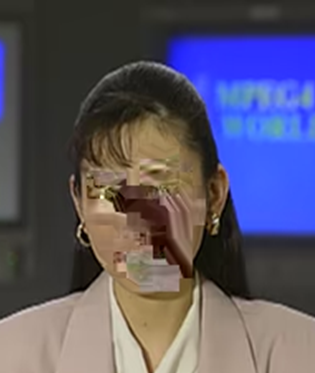}}
\hfill
\subfloat[Frame \#50\\Original]{
\includegraphics[width=0.110\textwidth]{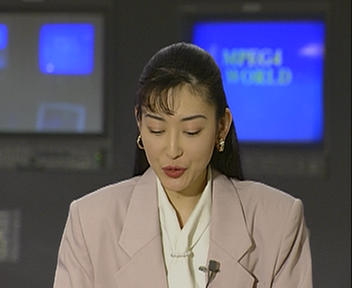}}
\hfill
\subfloat[Frame \#50\\Taha~\cite{taha2018end}]{
\includegraphics[width=0.110\textwidth]{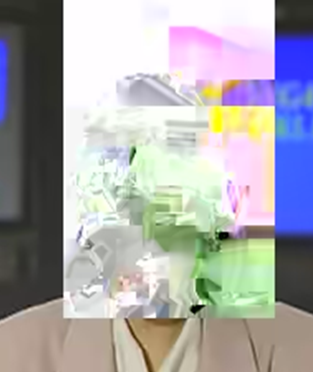}}
\hfill
\subfloat[Frame \#50\\Zhang~\cite{zhang2025visual}]{
\includegraphics[width=0.110\textwidth]{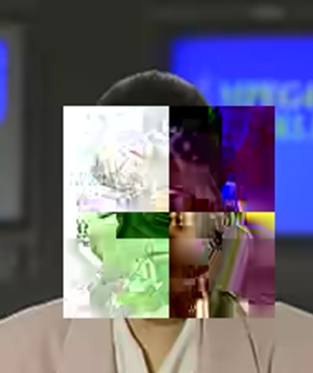}}
\hfill
\subfloat[Frame \#50\\Ours]{
\includegraphics[width=0.110\textwidth]{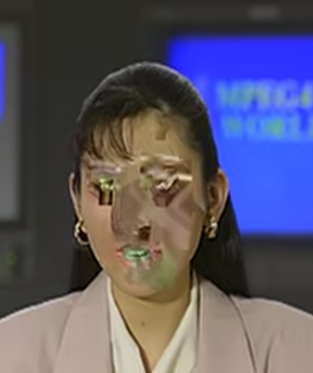}} \\[2pt]

\subfloat[Frame \#1\\Original]{
\includegraphics[width=0.110\textwidth]{figure/johnny_1ori.png}}
\hfill
\subfloat[Frame \#1\\Taha~\cite{taha2018end}]{
\includegraphics[width=0.110\textwidth]{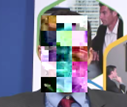}}
\hfill
\subfloat[Frame \#1\\Zhang~\cite{zhang2025visual}]{
\includegraphics[width=0.110\textwidth]{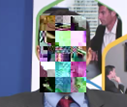}}
\hfill
\subfloat[Frame \#1\\Ours]{
\includegraphics[width=0.110\textwidth]{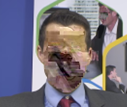}}
\hfill
\subfloat[Frame \#50\\Original]{
\includegraphics[width=0.110\textwidth]{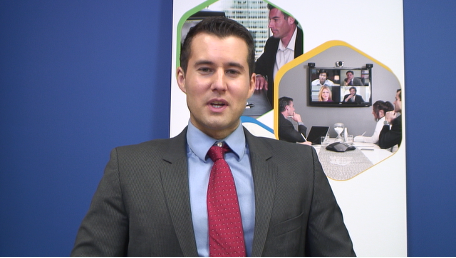}}
\hfill
\subfloat[Frame \#50\\Taha~\cite{taha2018end}]{
\includegraphics[width=0.110\textwidth]{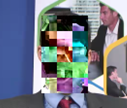}}
\hfill
\subfloat[Frame \#50\\Zhang~\cite{zhang2025visual}]{
\includegraphics[width=0.110\textwidth]{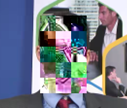}}
\hfill
\subfloat[Frame \#50\\Ours]{
\includegraphics[width=0.110\textwidth]{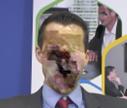}} \\[2pt]

\subfloat[Frame \#1\\Original]{
\includegraphics[width=0.110\textwidth]{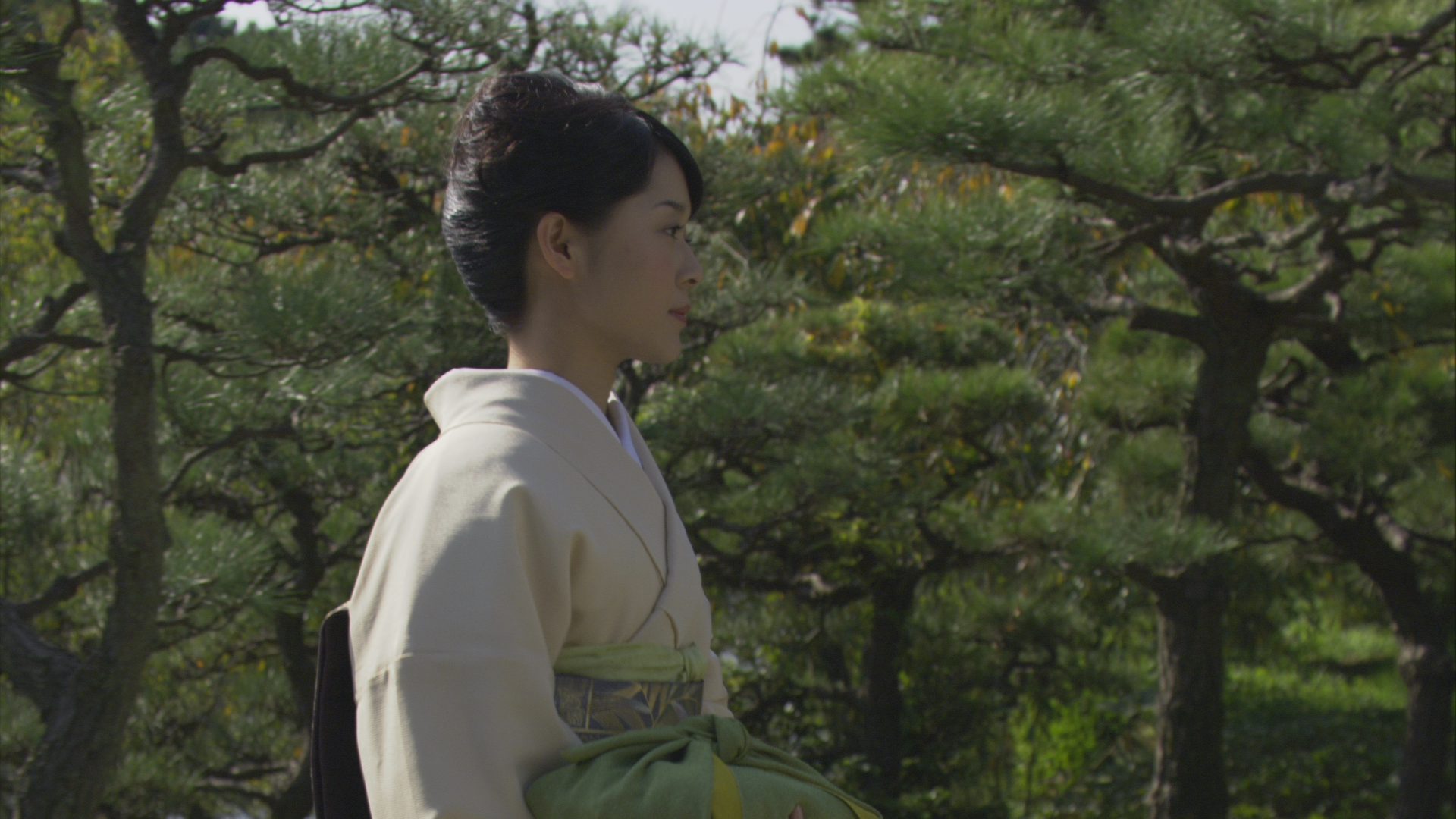}}
\hfill
\subfloat[Frame \#1\\Taha~\cite{taha2018end}]{
\includegraphics[width=0.110\textwidth]{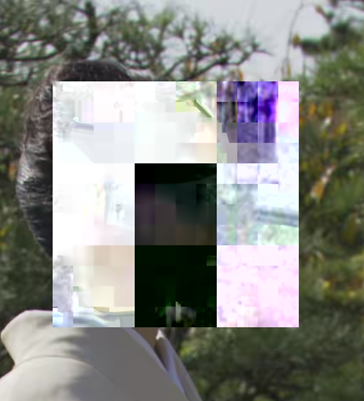}}
\hfill
\subfloat[Frame \#1\\Zhang~\cite{zhang2025visual}]{
\includegraphics[width=0.110\textwidth]{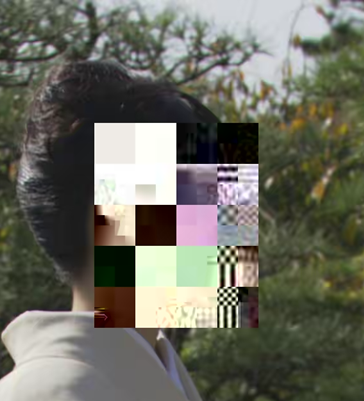}}
\hfill
\subfloat[Frame \#1\\Ours]{
\includegraphics[width=0.110\textwidth]{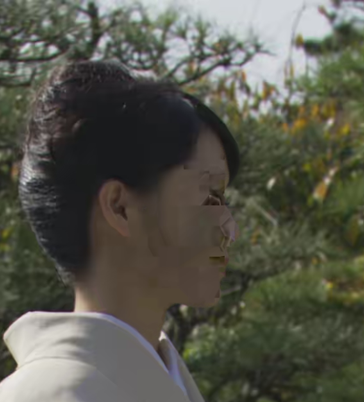}}
\hfill
\subfloat[Frame \#50\\Original]{
\includegraphics[width=0.110\textwidth]{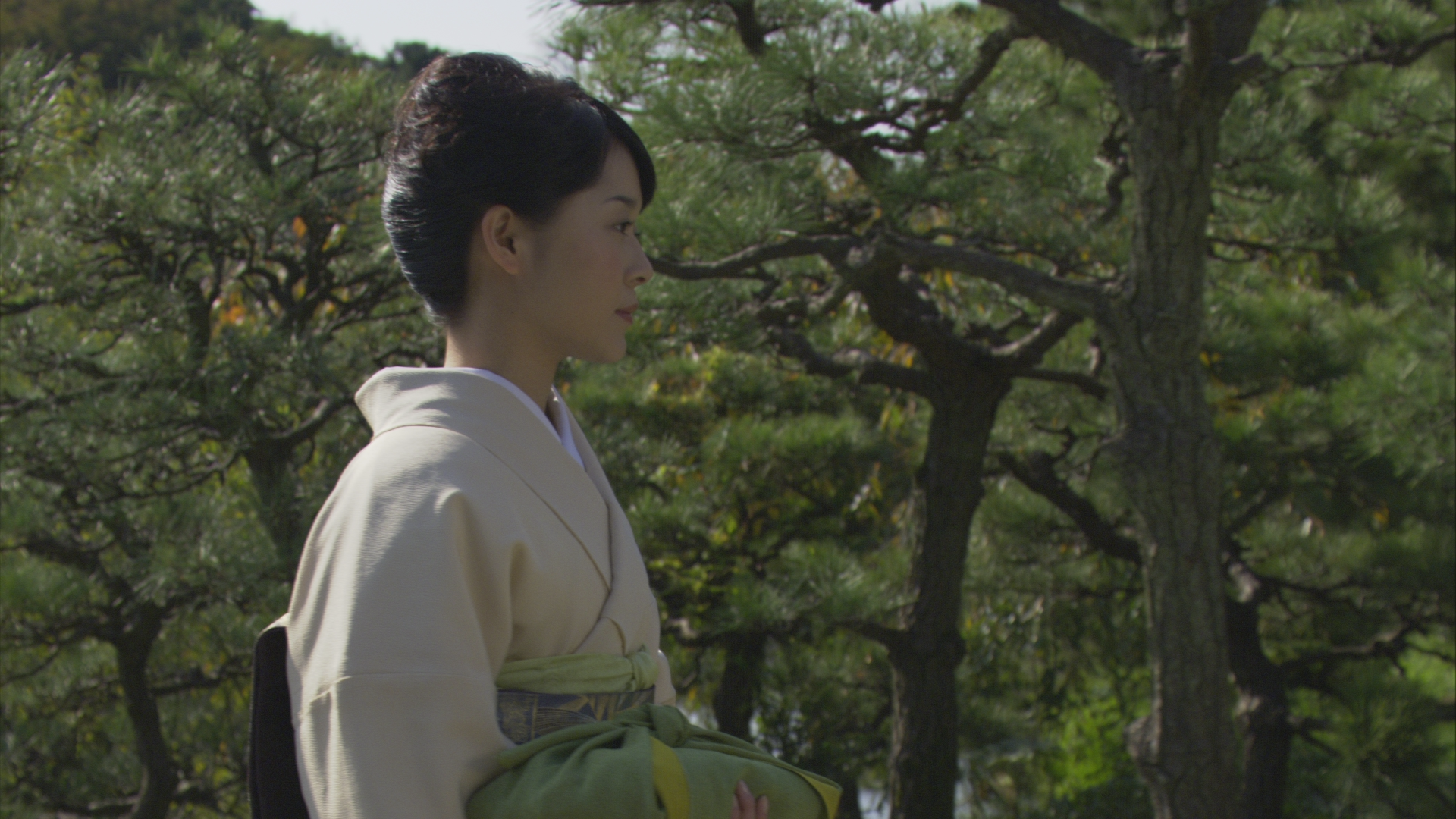}}
\hfill
\subfloat[Frame \#50\\Taha~\cite{taha2018end}]{
\includegraphics[width=0.110\textwidth]{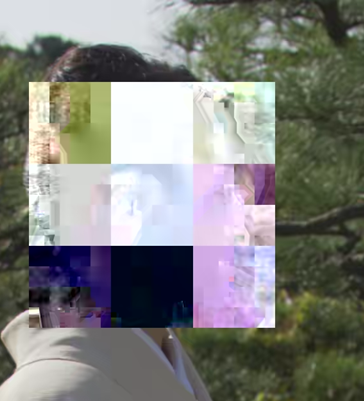}}
\hfill
\subfloat[Frame \#50\\Zhang~\cite{zhang2025visual}]{
\includegraphics[width=0.110\textwidth]{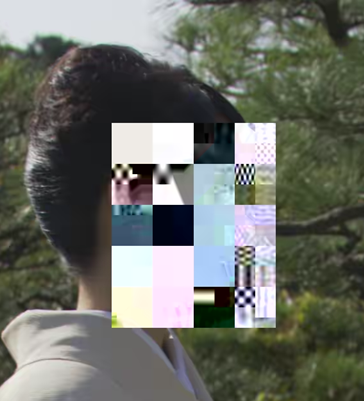}}
\hfill
\subfloat[Frame \#50\\Ours]{
\includegraphics[width=0.110\textwidth]{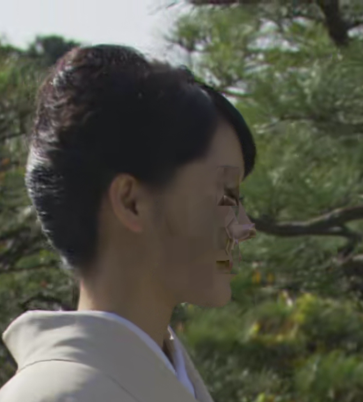}} \\[2pt]


\subfloat[Frame \#1\\Original]{
\includegraphics[width=0.110\textwidth]{figure/vidyo4-1-ori.png}}
\hfill
\subfloat[Frame \#1\\Taha~\cite{taha2018end}]{
\includegraphics[width=0.110\textwidth]{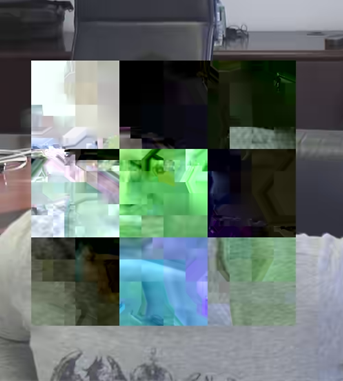}}
\hfill
\subfloat[Frame \#1\\Zhang~\cite{zhang2025visual}]{
\includegraphics[width=0.110\textwidth]{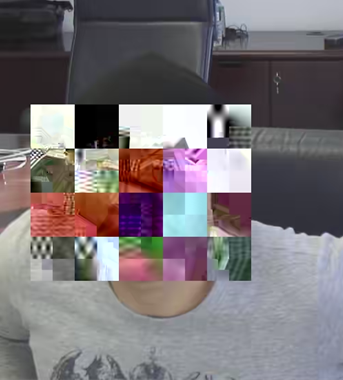}}
\hfill
\subfloat[Frame \#1\\Ours]{
\includegraphics[width=0.110\textwidth]{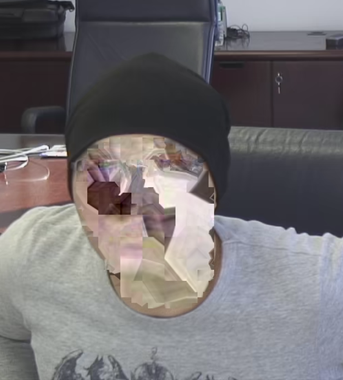}}
\hfill
\subfloat[Frame \#50\\Original]{
\includegraphics[width=0.110\textwidth]{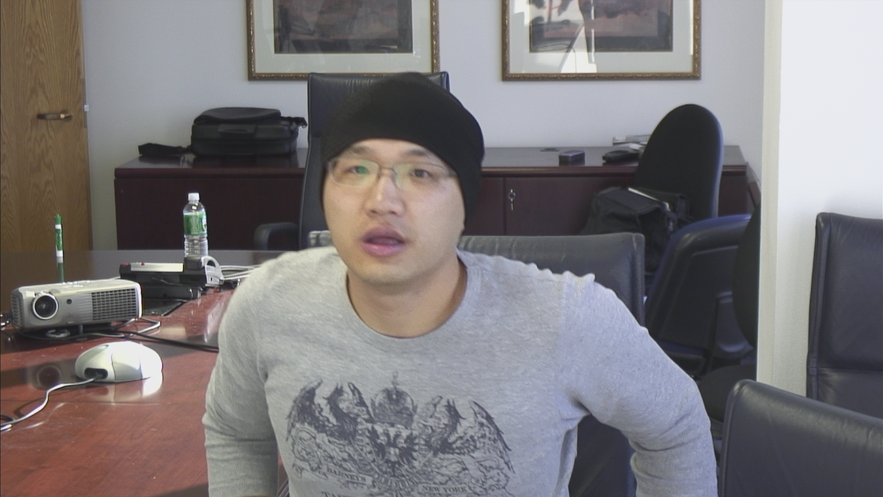}}
\hfill
\subfloat[Frame \#50\\Taha~\cite{taha2018end}]{
\includegraphics[width=0.110\textwidth]{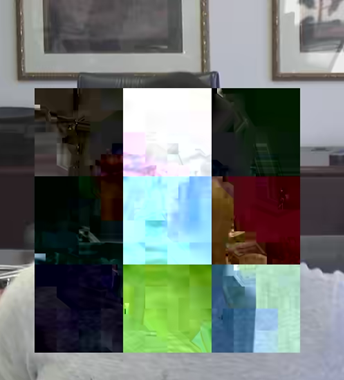}}
\hfill
\subfloat[Frame \#50\\Zhang~\cite{zhang2025visual}]{
\includegraphics[width=0.110\textwidth]{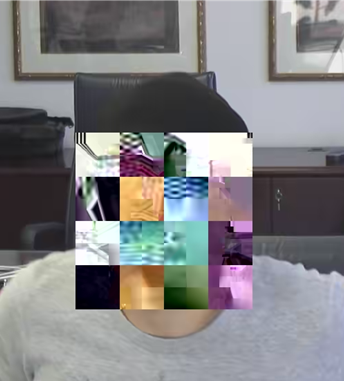}}
\hfill
\subfloat[Frame \#50\\Ours]{
\includegraphics[width=0.110\textwidth]{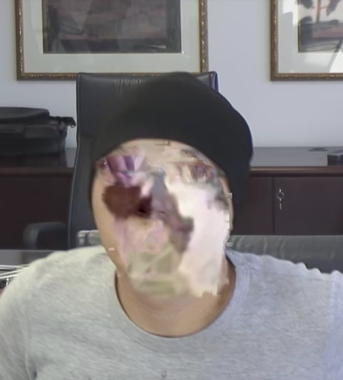}}

\caption{Visual comparison of ROI encryption results on four test sequences. Each row corresponds to one sequence: the original frame, the results encrypted by Taha~\cite{taha2018end}, Zhang~\cite{zhang2025visual}, and the proposed scheme for two representative frames. The proposed CU-level ROI encryption scheme achieves finer-grained protection and more accurate encryption localization than existing Tile-level approaches.}
\label{fig:detail}
\end{figure*}

1) \textit{Comparison of Subjective Vision}:
Figure \ref{fig:detail} shows a comparison of the perturbation effects on the different frames of the four video sequences using three comparative ROI encryption schemes at QP=24. As observed in the figure, Tile-level ROI encryption schemes \cite{taha2018end,zhang2025visual} are constrained by the fixed large rectangular blocks inherent in Tile coding. This lack of flexibility in encryption granularity leads to significant encryption redundancy. when an ROI lies on a Tile boundary or occupies only a small portion of a Tile, the algorithm must encrypt the ROI using multiple Tiles to ensure complete privacy masking. This results in extensive ineffective encryption of non-sensitive background pixels surrounding the ROI, degrading the video's visual quality. However, due to H.265/HEVC's recursive quadtree partitioning mechanism, the CU can adaptively scale based on the geometric characteristics of the ROI contour. This enables ROI encryption boundaries to be precise up to 8$\times$8-sized CU. Consequently, the CU-level encryption scheme demonstrates superior boundary fitting capabilities. It can align with the natural contours of the ROI at finer increments, effectively preserving the background whiteboard content. This ensures that while protecting the privacy of the ROI, viewers can still obtain critical contextual information. It is suitable for application scenarios requiring a balance between privacy security and behavioral analysis.

In addition, to further evaluate the visual quality of the encrypted videos, we conduct a subjective assessment based on the Differential Mean Opinion Score (DMOS) and invited 120 participants from diverse professional backgrounds to subjectively evaluate the visual differences between each group of frames. A five-level rating scale was adopted in Table \ref{DMOS}, where higher scores indicate better visual perturbation effects. Specifically, question 1 describes the degree of disturbance to the facial features, question 2 describes whether the proposed scheme is superior to Tile-level encryption schemes. The resulting DMOS scores are summarized in Table \ref{DMOS-score}.

\begin{table}[htbp]
\centering
\footnotesize
\caption{Five-Level subjective rating scale used
in the DMOS evaluation}
\label{DMOS}
\centering 
\setlength{\tabcolsep}{2pt}
\begin{tabular}{ccc} 
\hline
\hline
\multirow{2}{*}{Score} & \multicolumn{2}{c}{Description} \\ 
\cline{2-3} 
 & Question 1 & Question 2\\ 
\hline
1 & Almost undisturbed  & Clearly inferior to Tile-level schemes   \\
2 & Minor disturbance & Slightly inferior to Tile-level schemes   \\
3 & Disturbed to a certain extent & Equivalent to Tile-level schemes  \\
4 & Obvious disturbance & Slightly better than Tile-level schemes  \\
5 & Severe disturbance & Clearly superior to Tile-level schemes   \\ 
\hline
\hline
\end{tabular}
\label{DMOS}
\end{table}

\begin{table}[htbp]
\centering
\footnotesize
\caption{Average DMOS scores for different sequences}
\label{DMOS-score}
\centering 
\setlength{\tabcolsep}{12pt}
\begin{tabular}{ccc} 
\hline
\hline
\multirow{2}{*}{Sequence} & \multicolumn{2}{c}{Average DMOS Score} \\ 
\cline{2-3} 
 & Question 1 & Question 2\\ 
\hline
Foreman & 4.725  & 4.458   \\
Akiyo & 4.267 & 4.300   \\
Vidyo3 & 4.167 & 4.450  \\
Vidyo4 & 4.617 & 4.500  \\ 
\hline
\hline
\end{tabular}
\label{DMOS-score}
\end{table}
As shown in the results, the average score of most sequences is above 4.3, demonstrating that, from a subjective visual perspective, most participants believe that the proposed scheme can fully protect ROI region and consider it superior to existing schemes.

2) \textit{Comparison of Objective Indicators}:
Zhang et al.\cite{zhang2025visual} summarized the typical objective indicators and their calculation formulas and evaluation dimensions used in ROI selective encryption, including PSNR, SSIM, EDR, information entropy, NPCR, UACI, and bitrate changes. Table \ref{obj_indicators} shows the average values of various indicators for the five video sequences under five encryption schemes \cite{sheng2024chaos,sheng2024contentt,sheng2024contents,zhang2025visual,taha2018end}. It is worth noting that, in addition to the two ROI encryption schemes, we further introduce three recent full-frame encryption schemes to better demonstrate the distortion effects of different schemes on the objective metrics. For the full-frame encryption schemes, the results are calculated over the entire frame.

\begin{table}[htbp]
\caption{Comparison of objective indicators of different encryption schemes}
\centering
\footnotesize
\renewcommand{\arraystretch}{1.2}
\setlength{\tabcolsep}{3pt}
\resizebox{\columnwidth}{!}{%
\begin{tabular}{cccc|ccc}
\hline \hline
\multirow{2}{*}{Objective Indicators}  &
\multicolumn{3}{c|}{Full-frame Encryption} &
\multicolumn{3}{c}{ROI Encryption} \\
\cline{2-4}\cline{5-7}
 & 
Sheng \cite{sheng2024chaos} &
Fu \cite{sheng2024contentt} &
Tie \cite{sheng2024contents} &
Taha \cite{taha2018end} &
Zhang \cite{zhang2025visual} &
Proposed \\
\hline

PSNR
 & 10.65 & 10.45 & 11.36 & 12.94 & 12.21 & \cellcolor{myblue} 17.29 \\

SSIM
 & 0.262 & 0.260 & 0.308 & 0.335 & 0.278 & \cellcolor{myblue} 0.485 \\

EDR
 & 0.938 & 0.885 & 0.906 & 0.870 & 0.921 & \cellcolor{myblue} 0.772 \\

Infomation Entropy
 & 7.39 & 7.38 & 7.41 & 7.58 & 7.65 & \cellcolor{myblue} 7.11 \\

NPCR
 & 99.51 & 99.61 & 99.57 & 99.48 & 99.55 & \cellcolor{myblue} 89.43 \\

UACI
 & 35.62 & 33.91 & 33.92 & 26.51 & 27.04 & \cellcolor{myblue} 22.03 \\

Bitrate Change
 & 2.96 & 1.92 & 3.00 & 1.65 & 9.05 & \cellcolor{myblue} 13.70 \\

\hline \hline
\end{tabular}}
\label{obj_indicators}
\end{table}

The experimental results show that the objective indicator results of the proposed CU-level ROI encryption schemes are weaker than those of existing Tile-level ROI encryption schemes and full-frame encryption schemes. This phenomenon is caused by the inherent mechanism of fine-grained encryption, and a detailed theoretical analysis is provided in section \ref{anly}. In fact, both the objective metrics and the subjective visual results clearly demonstrate that the proposed scheme introduces sufficient encryption distortion to effectively protect the security of ROI regions.

\subsection{Analysis of Distortion Difference Between Tile-level and CU-level Encryption}\label{anly}
For identical facial regions within the same video sequence, experimental results indicate that while CU-level encryption achieves more precise spatial alignment with facial contours, the resulting distortion intensity is generally weaker than that of Tile-level encryption. This phenomenon stems not from the encryption algorithms themselves, but primarily from differences in ROI spatial coverage and encoding structure isolation mechanisms. This section provides a detailed analysis of this phenomenon.

\subsubsection{ROI Spatial Coverage Difference Between Tile-level and CU-level Encryption}\label{sub1}
In Tile-level ROI encryption, the encrypted area uses fixed-size Tiles as the fundamental unit. Due to the irregular shape and fine boundaries of the facial regions, a single Tile often contains complex and diverse elements such as the face, hair, clothing, and portions of the background. In contrast, CU-level ROI encryption can more precisely align with facial contours, concentrating the encrypted area primarily on the semantically consistent, relatively simple-textured facial region.

To characterize the complexity of the content within the encrypted region, this paper introduces the residual energy per pixel as a metric. Let $\Omega$ denote a region of interest (ROI), whose prediction residual in pixel-domain is defined as:
\begin{equation}e(x,y)=I(x,y)-p(x,y),\quad(x,y)\in\Omega\end{equation}
where $I(x,y)$ represents the pixel value of the original frame, $p(x,y)$ denotes the predicted pixel value, and $e(x,y)$ signifies the prediction error. The residual energy per pixel in this region can be expressed as:
\begin{equation}E(\Omega)=\frac{1}{|\Omega|}\sum_{(x,y)\in\Omega}[e(x,y)]^2\label{equ7}\end{equation}
where $|\Omega|$ denotes the number of pixels within the ROI.

To illustrate the relationship between residual energy and coding distortion, this paper models coding perturbations using the Parseval's Theorem \cite{pfister2017discrete}. The Parseval theorem states that the energy of a signal in the time domain is equivalent to its energy in the frequency domain. In H.265/HEVC video coding, this manifests as the error energy in the pixel domain being equivalent to the error energy in the DCT transform coefficient domain. Here, we illustrate encryption distortion using a  bit flipping encryption strategy. Let the transform coefficient be $c(u,v)$. For a symbol bit flipping encryption strategy, the sign of the transform domain coefficient $c(u,v)$ is inverted, yielding the encrypted coefficient $-c(u,v)$. The perturbation coefficient difference is $2c(u,v)$. Therefore, the perturbation energy is:
\begin{equation}|\Delta c(u,v)|^2=|2c(u,v)|^2=4|c(u,v)|^2\end{equation}

By Parseval's Theorem:
\begin{equation}\sum_{(x,y)\in\Omega}|\Delta e(x,y)|^2=\sum_{(u,v)}|\Delta c(u,v)|^2=4\sum_{(u,v)}|c(u,v)|^2\label{equ9}\end{equation}
where $\Delta e(x,y)$ denotes the error in the pixel domain introduced by encryption, while $\Delta c(u,v)$ represents the error in the transform domain coefficient introduced by encryption.

By the DCT energy conservation principle, the total energy of the original residuals in the pixel domain equals the total energy of the original coefficients in the transform domain, yielding the following:
\begin{equation}\sum_{(x,y)\in\Omega}[e(x,y)]^2=\sum_{(u,v)}|c(u,v)|^2\label{equ10}\end{equation}

Replace $\sum_{(u,v)}|c(u,v)|^2$ in Equation (\ref{equ9}) with $\sum_{(x,y)\in\Omega}[e(x,y)]^2$ in Equation (\ref{equ10}), we obtain:
\begin{equation}\sum_{(x,y)\in\Omega}|\Delta e(x,y)|^2=4\sum_{(x,y)\in\Omega}[e(x,y)]^2\label{equ11}\end{equation}

Substitute Equation (\ref{equ7}) into Equation (\ref{equ11}), we obtain:
\begin{equation}\sum_{(x,y)\in\Omega}|\Delta e(x,y)|^2=4\cdot|\Omega|\cdot E(\Omega)\label{equ12}\end{equation}

Define the average energy error within the encrypted region as:
\begin{equation}D(\Omega)=\frac{1}{|\Omega|}\sum_{(x,y)\in\Omega}|\Delta e(x,y)|^2\label{equ13}\end{equation}

Through Equation (\ref{equ12}) and (\ref{equ13}), we obtain:
\begin{equation}D(\Omega)=4\cdot E(\Omega)\label{14}\end{equation}

It follows that, under the same encryption strategy, the average energy error $D(\Omega)$ introduced by encryption is proportional to the residual energy $E(\Omega)$ of the region itself. We then select five video sequences from Table \ref{tab2} and calculate the average residual energy $E(\Omega)$ within the ROI for the first 100 frames of each sequence. It can be seen from the results, which are shown in Table \ref{tab:residual}, the average residual energy $E(\Omega)$ of Tile-level ROI is much larger than that of CU-level ROI, we can thus derive that:
\begin{equation}E(\Omega_{\text{Tile}})>E(\Omega_{\text{CU}})\label{15}\end{equation}
where $\Omega_{\text{Tile}}$ and $\Omega_{\text{CU}}$ are the Tile-level ROI and CU-level ROI, respectively. Substituting Equation (\ref{15}) into Equation (\ref{14}), we obtain:
\begin{equation}D(\Omega_{\text{Tile}})>D(\Omega_{\text{CU}})\label{16}\end{equation}

Therefore, under the same encryption strategy, The larger the residual energy of a region, the greater its average error energy, that is, the stronger the distortion generated after encryption. Since the Tile-level ROI usually covers more heterogeneous content and exhibits higher residual energy than the CU-level ROI, it consequently produces larger average distortion after encryption.

\begin{table}[htbp]
\centering
\footnotesize
\caption{\small Comparison of the average residual energy $E(\Omega)$ within different ROI regions}
\label{tab:erd_values}
\setlength{\tabcolsep}{4pt}
\begin{tabular}{lcccccc} 
\hline
\hline
\multirow{2}{*}{Sequence} & \multicolumn{6}{c}{$E(\Omega)$} \\ 
\cline{2-7} 
 & Akiyo & PartyScene & Johnny & Vidyo4 & Kimono & Average\\ 
\hline
Tile-level ROI & 77.91 & 18.98 & 27.29 & 26.66 & 9.97 & 32.16\\
CU-level ROI & 68.47 &  12.93 &  11.19 & 17.96 & 2.35 & 22.58\\

\hline
\hline
\end{tabular}
\label{tab:residual}
\end{table}

\subsubsection{Encoding Structure Isolation Difference Between Tile-level and CU-level Encryption}\label{SUB2}
Beyond differences in spatial coverage of ROI, the distinct isolation in encoding structures between Tile-level and CU-level encryption are also primary causes of varying distortion performance. In the H.265/HEVC standard, the reconstructed pixel value of a CU depends on both the residual information of the current block and the predictive information from the reference block. When encrypting syntax elements in the compressed domain, the final reconstruction distortion results from two sources: distortion directly caused by encrypting the syntax elements within the current block, and propagated distortion transmitted from the reference block.

Assuming $L\in\{\mathrm{Tile,CU}\}$ represents two different encryption strategies, the total distortion of the $n$-th CU under encryption strategy $L$ can be expressed as:
\begin{equation}D_{\mathrm{total}}^L(n)=D_{\mathrm{enc}}^L(n)+\lambda\cdot\gamma_L\cdot D_{\mathrm{total}}^L(n-1),L\in\{\mathrm{Tile,CU}\}\label{equ1}\end{equation}
where $D_{\mathrm{enc}}^L(n)$ represents the distortion caused by encrypting syntax elements of the current $n$-th CU under encryption strategy $L$, $D_{\mathrm{total}}^L(n-1)$ represents the total distortion of the previous $n-1$ CU along the prediction path under encryption strategy $L$, $\gamma_L\in[0,1]$ represents the proportion of ROI in the predicted information of the current CU under encryption strategy $L$, while $\lambda$ is the propagation coefficient describing the attenuation of error transfer from the reference CU to the current CU. If $\lambda\geq1$, it implies error is transferred losslessly to the next level or amplified infinitely during prediction, leading to decoding image collapse and violating coding stability requirements. Therefore, $\lambda\in[0,1)$.

We use the same syntax elements for encryption at Tile-level and CU-level, In order to clarify the differences in encoding structure isolation mechanisms between Tile-level and CU-level, it is assumed that the distortion caused by syntax elements is consistent for each CU, that is:
\begin{equation}
    D_{\mathrm{enc}}^L(n)=D_{\mathrm{enc}}^L(n-1)=D_{\mathrm{enc}}^L(n-2)=\cdots=D_{\mathrm{enc}}^L(0)=\alpha
\end{equation}

The result of expanding Equation (\ref{equ1}) after one recursive step is:

\begin{equation}\begin{aligned}
D_{\mathrm{total}}^L(n) =\alpha+\lambda\cdot\gamma_L\cdot D_{\mathrm{total}}^L(n-1) \\=
  \alpha+(\lambda\cdot\gamma_L)[\alpha+(\lambda\cdot\gamma_L)D_{\mathrm{total}}^L(n-2)]
\end{aligned}\end{equation}

The final result when recursively calling $n=0$ is:
\begin{equation}\label{eq20}
D_\mathrm{total}^L(n)=\alpha\sum_{k=0}^{n-1}(\lambda\cdot\gamma_L)^k+(\lambda\cdot\gamma_L)^nD_\mathrm{total}^L(0)
\end{equation}

To facilitate a subsequent comparison of the distortion levels of different encryption strategies, we use the sum of a geometric series to evaluate $\sum_{k=0}^{n-1}(\lambda\cdot\gamma_L)^k$, yielding:
\begin{equation}\label{eq21}
    \sum_{k=0}^{n-1}(\lambda\cdot\gamma_L)^k=\frac{1-(\lambda\cdot\gamma_L)^n}{1-(\lambda\cdot\gamma_L)},(\lambda\cdot\gamma_L)\neq1
\end{equation}

Therefore, Substituting Equation (\ref{eq21}) into Equation (\ref{eq20}), we get:
\begin{equation}D_{\mathrm{total}}^L(n)=\alpha\frac{1-(\lambda\cdot\gamma_L)^n}{1-(\lambda\cdot\gamma_L)}+(\lambda\cdot\gamma_L)^nD_{\mathrm{total}}^L(0)\end{equation}

Although the number of CUs propagating along the prediction chain is limited in the actual encoding process, when the error propagation path of the encryption process is long enough, the distortion differences of different encryption strategies can be more clearly explained by calculating the limit of $D_{\mathrm{total}}^L(n)$. Since $\gamma_L\in[0,1]$ and $\lambda\in[0,1)$, we obtain $(\lambda\cdot\gamma_{L})\in[0,1)$. Then, the limit $D_{\mathrm{total}}^L(n)$ can be expressed as:
\begin{equation}\begin{aligned}
\lim_{n\to\infty}D_{\mathrm{total}}^L(n)=
\frac{\alpha}{1-(\lambda\cdot\gamma_L)}\end{aligned}\end{equation}

In Tile-level ROI encryption, due to the encoding structure isolation constraint at Tile boundaries, encoding blocks within a Tile can only reference reconstruction results from the same Tile during prediction, resulting in $\gamma_{\text{Tile}}=1$. In CU-level encryption, however, since CUs can mutually reference neighborhood information, CUs at ROI boundaries can reference undisturbed CUs outside the ROI, leading to $\gamma_{\text{CU}}<1$.

Since $\lambda\in[0,1)$, then:
\begin{equation}\frac{\alpha}{1-(\lambda\cdot\gamma_{\text{Tile}})}>\frac{\alpha}{1-(\lambda\cdot\gamma_{\text{CU}})}\end{equation}

Ultimately:
\begin{equation}
\lim_{n\to\infty}D_{\mathrm{total}}^{\text{Tile}}(n)>\lim_{n\to\infty}D_{\mathrm{total}}^{\text{CU}}(n)
\end{equation}

Therefore, when the encryption space is large enough, Tile-level encryption can accumulate stronger distortion. In summary, based on the derivations in Sections \ref{sub1} and \ref{SUB2}, it can be observed that both the ROI spatial coverage and encoding structure isolation mechanisms lead to stronger perturbation effects in tile-level encryption than in CU-level encryption.

\section{Conclusions}\label{sec6}
This article proposes an H.265/HEVC fine-grained ROI video encryption algorithm based on encoding units and hint segmentation. In response to the problems of over encryption and insufficient encryption in traditional Tile level ROI encryption schemes, a more adaptive CU level ROI encryption strategy is designed to provide finer contour control for the ROI encryption scheme. In response to encryption diffusion, this article proposes the use of PCM encoding mode as a boundary isolation mechanism, effectively blocking pixel dependent propagation between ROI and non-ROI, thereby improving the independence and controllability of region encryption. The experimental results show that the algorithm can achieve more detailed regional perturbation effects while ensuring the safety of the ROI. However, there is still room for improvement in terms of disturbance effects and solving diffusion problems in this article. future work will attempt to introduce a combination of chaos models and coefficient perturbation mechanisms to enhance the randomness and unpredictability of ciphertext, and improve the robustness of regional perturbations; In solving the problem of diffusion phenomenon, we can try to refer to Tile's isolation idea and seek a solution to cut off cross domain references between CUs, in order to improve disturbance strength while controlling the increase of bit rate.

\bibliographystyle{ieeetr}
\bibliography{reference}

\end{document}